\documentstyle[preprint,aps,pictex]{revtex}

\newcommand{\be}{\begin{equation}}
\newcommand{\ee}{\end{equation}}
\newcommand{\bea}{\begin{eqnarray}}
\newcommand{\eea}{\end{eqnarray}}
\newcommand{\ba}{\begin{array}}
\newcommand{\ea}{\end{array}}

\def\npb#1#2#3{Nucl.\ Phys.\ {\bf B#1}, #2 (#3)}
\def\prd#1#2#3{Phys.\ Rev.\ {\bf D#1}, #2 (#3)}
\def\plb#1#2#3{Phys.\ Lett.\ {\bf B#1}, #2 (#3)}
\def\zpc#1#2#3{Z.\ Phys. {\bf C#1}, #2 (#3)}

\begin{document}

\preprint{\begin{tabular}{r} KAIST--TH 97/20\\[-3mm] FTUAM 97/16 \end{tabular}}

\title{Four-Dimensional Effective Supergravity and Soft Terms in $M$--Theory}
\author{Kiwoon Choi$^1$, Hang Bae Kim$^1$ and Carlos Mu\~noz$^{1,2}$}
\address{
$^1$Department of Physics,
Korea Advanced Institute of
Science and Technology \\
Taejon 305-701, Korea \\
$^2$Departamento de F{\'\i}sica Te\'orica C-XI,
Universidad Aut\'onoma de Madrid \\
Cantoblanco, 28049 Madrid, Spain}

\maketitle

\begin{abstract}
We provide a simple macroscopic analysis of the four-dimensional effective
supergravity of the Ho\v{r}ava-Witten $M$-theory which is expanded
in powers of $\kappa^{2/3}/\rho V^{1/3}$ and $\kappa^{2/3}\rho/V^{2/3}$
where $\kappa^2$, $V$ and $\rho$ denote the eleven-dimensional gravitational
coupling, the Calabi-Yau volume and the eleventh length respectively.
Possible higher order terms in the K\"ahler potential are identified and
matched with the heterotic string corrections.  In the context of this
$M$-theory expansion, we analyze the soft supersymmetry--breaking terms
under the assumption that supersymmetry is spontaneously broken by the
auxiliary components of the bulk moduli superfields.  It is examined how
the pattern of soft terms changes when one moves from the weakly coupled
heterotic string limit to the $M$-theory limit.
\end{abstract}

\pacs{}


Recently Ho\v{r}ava and Witten proposed that the strong coupling limit
of $E_8\times E_8$ heterotic string theory can be described by
11-dimensional supergravity (SUGRA) on a manifold with boundary where
the two $E_8$ gauge multiplets are restricted to the two 10-dimensional
boundaries respectively ($M$-theory) \cite{Horava-Witten}.
The effective action of this  limit has been  systematically analyzed
in an expansion in powers of $\kappa^{2/3}$ where $\kappa^2$ denotes
the 11-dimensional gravitational coupling.  At zeroth order, the
effective action is simply that of the 11-dimensional supergravity
which includes only the bulk fields whose kinetic terms are of order
$\kappa^{-2}$.  However at first order in the expansion, there appear
a variety of additional terms including the 10-dimensional boundary
action of the $E_8\times E_8$ gauge multiplets which is of order
$\kappa^{-4/3}$.  It was also noted that a four-gaugino term appears
at higher order which would lead to an interesting phenomenological
consequence when supersymmetry (SUSY) is broken by the gaugino
condensation on the hidden boundary \cite{Horava}.

Some phenomenological implications of the strong-coupling limit of
$E_8\times E_8$  heterotic string theory has been studied by
compactifying the 11-dimensional $M$-theory on a Calabi-Yau manifold
times the eleventh segment \cite{Witten}.  The resulting 4-dimensional
effective theory can reconcile the observed Planck scale
$M_{P}=1/\sqrt{8\pi G_N}\approx2.4\times 10^{18}$ GeV with the
phenomenologically favored GUT scale $M_{GUT}\approx3\times 10^{16}$ GeV
in a natural manner, providing an attractive framework for the
unification of couplings \cite{Witten,Banks-Dine}.  In this framework,
$M_{GUT}$ corresponds to $1/V^{1/6}$ where $V$ is the Calabi-Yau volume,
while $M_{P}^2=2\pi\rho V/\kappa^2$ for $\pi\rho$ denoting the length of
the eleventh segment.  Then by choosing $\pi\rho\approx4V^{1/6}$ together
with $M_{GUT}=1/V^{1/6}\approx 3\times 10^{16}$ GeV, one could get the
correct values of $\alpha_{GUT}$ and $M_P$, which was not allowed in the
weakly coupled heterotic string theory.  An additional phenomenological
virtue of the $M$-theory limit is that there can be a QCD axion whose
high energy axion potential is suppressed enough so that the strong CP
problem can be solved by the axion mechanism \cite{Banks-Dine,Choi}.
These phenomenological virtues have motivated many of the recent studies
of the 4-dimensional effective theory of $M$-theory
\cite{Caceres-Kaplunovsky-Mandelberg,Mourad,Li-Lopez-Nanopoulos1,%
Dudas-Grojean,Antoniadis-Quiros1,Nilles-Olechowski-Yamaguchi,%
Lalak-Thomas,Dudas,Ellis-Faraggi-Nanopoulos,Lukas-Ovrut-Waldram}.

As is well known, the 4-dimensional effective action of the weakly
coupled heterotic string theory can be expanded in powers of the
two dimensionless variables: the string coupling
$\epsilon_s=e^{2\phi}/(2\pi)^5$ 
and the worldsheet sigma model coupling
$\epsilon_{\sigma}=4\pi\alpha^{\prime}/V^{1/3}$
\cite{Green-Schwarz-Witten}.
The effective action of $M$-theory can be similarly analyzed 
by expanding it in powers of the two dimensionless variables:
$\epsilon_1=\kappa^{2/3}\pi\rho/V^{2/3}$ and $\epsilon_2=
\kappa^{2/3}/\pi\rho V^{1/3}$.
To compute the 4-dimensional effective action,
one first expands  the 11-dimensional action in powers
of $\kappa^{2/3}$ to find the compactification solution which is 
expanded in powers of $\epsilon_1$ and $\epsilon_2$.
The subsequent Kaluza-Klein reduction of the 11-dimensional action
for this  compactification solution 
will lead to the desired 4-dimensional effective action.
At the leading order in this expansion,
the K\"ahler potential, superpotential and gauge kinetic functions
have been computed in
\cite{Banks-Dine,Li-Lopez-Nanopoulos1,Dudas-Grojean,%
Nilles-Olechowski-Yamaguchi}.
It is rather easy to determine 
the order $\epsilon_1$ correction to the leading
order gauge kinetic functions
\cite{Banks-Dine,Choi,Nilles-Olechowski-Yamaguchi,Nilles-Stieberger},
while it is much more nontrivial to compute
the order $\epsilon_1$ correction to
the leading order K\"ahler potential,
which was recently done by  Lukas, Ovrut and Waldram 
\cite{Lukas-Ovrut-Waldram}.
As we will argue later,
the holomorphy and Peccei--Quinn symmetries guarantee
that there is {\it no} further correction
to the gauge kinetic functions and the superpotential 
at any finite order
in the $M$-theory expansion \cite{Nilles}. 
However generically the K\"ahler potential is expected to 
receive corrections which are higher order in $\epsilon_1$ or 
$\epsilon_2$.
An explicit computation of these  higher order corrections
will be highly nontrivial  
since first of all  the 11-dimensional action is known
only up to the terms of order $\kappa^{2/3}$ relative to the zeroth
order action (except for the order $\kappa^{4/3}$ four-gaugino term)
and secondly the higher order computation of the compactification solution
and its  Kaluza-Klein reduction
are  much more complicated.

In this paper, we wish to provide a simple macroscopic analysis
of the 4-dimensional effective SUGRA action by expanding it 
in powers of $\epsilon_1$ and $\epsilon_2$, 
and apply the results for the computation of soft 
SUSY--breaking terms.  
As we will see, this analysis allows us to
extract the form of possible higher order corrections to the K\"ahler
potential and also to estimate their size for the physically
interesting values of moduli.
An interesting feature of the 4-dimensional effective SUGRA is that
the  forms of the gauge kinetic function and 
superpotential are {\it not} changed (up to nonperturbative corrections)
when one moves from the weakly coupled heterotic string domain 
to the $M$-theory domain in the moduli space.
The same is true for the leading order K\"ahler potential,
i.e. the K\"ahler potential of the weakly coupled heterotic string
computed at the leading  order in the string loop and sigma model
perturbation theory is the same as the $M$-theory K\"ahler potential
computed at the leading order in the $M$-theory expansion.
In fact, one can smoothly move from the $M$-theory domain 
with $\epsilon_s\gg 1$ to the heterotic
string domain with $\epsilon_s\ll 1$ 
while keeping $\epsilon_1$ and $\epsilon_2$ to be  small enough.
This means that the $M$-theory K\"ahler potential expanded in
$\epsilon_1$ and $\epsilon_2$ is valid even on the heterotic
string domain up to (nonperturbative) corrections which
can not be taken into account by the $M$-theory expansion.
Then as in the case of the gauge kinetic functions,
one can  determine the expansion coefficients 
in the $M$-theory K\"ahler potential 
by matching  the heterotic string K\"ahler potential
which can be computed in   
the string loop and sigma model perturbation theory.

About the issue of SUSY breaking, the possibility of SUSY breaking by the
gaugino condensation on the hidden boundary has been studied
\cite{Horava,Nilles-Olechowski-Yamaguchi,Lalak-Thomas}
and also some interesting features of the
resulting soft SUSY--breaking terms were discussed in
\cite{Nilles-Olechowski-Yamaguchi}.
Here we analyze the soft SUSY--breaking terms under the more general
assumption that SUSY is spontaneously broken by
the auxiliary components of the bulk moduli superfields in the model.
We examine in particular how the soft terms
vary when one moves from the weakly coupled   heterotic string
limit to the $M$-theory limit in order to see whether
these two limits can be distinguished  by the soft term physics.
Our analysis implies that there can be a sizable difference between
the heterotic string limit and the $M$-theory limit 
even in the overall pattern of soft terms.
 
Let  us first discuss  possible perturbative expansions
of the 4-dimensional effective SUGRA action of $M$-theory.
As in the case of weakly coupled heterotic string theory,
the effective SUGRA of compactified $M$-theory
contains two model--independent moduli
superfields $S$ and $T$ whose scalar components can be identified as 
\bea
{\rm Re}(S)&=&\frac{1}{2\pi}  (4\pi\kappa^2)^{-2/3}V\ ,
\nonumber \\
{\rm Re}(T)&=& \frac{6^{1/3}}{8\pi}(4\pi
\kappa^2)^{-1/3}\pi \rho V^{1/3}\ ,
\label{dilaton-modulus}
\eea
where these normalizations of $S$ and $T$ have been chosen to keep the
conventional form of the gauge kinetic functions in the effective SUGRA.
(See (\ref{gaugef}) for our form of the gauge kinetic functions.
Our $S$ and $T$ correspond to $\frac{1}{4\pi}S$ and $\frac{1}{8\pi}T$ 
of \cite{Choi} respectively.)
The pseudoscalar components ${\rm Im}(S)$ and ${\rm Im}(T)$
are the well known model--independent axion and the K\"ahler axion
whose couplings are constrained by the non--linear Peccei--Quinn
symmetries:
\be
U(1)_S: S\rightarrow S+i\beta_S\ ,
\quad
U(1)_T: T\rightarrow T+i\beta_T\ ,
\label{PQ-symmetry}
\ee
where $\beta_S$ and $\beta_T$ are continuous real numbers. 
These Peccei-Quinn symmetries  appear as exact
symmetries at any finite order in the $M$-theory expansion, 
but they are explicitly 
broken by nonperturbative effects that will not be taken into
account in this paper.

The moduli $S$ and $T$ can be used to define
various kind of  expansions which may be applied for
the low--energy effective action. For instance, in the weakly
coupled heterotic string limit,
we have
\bea
&&{\rm Re}(S)=e^{-2\phi}\frac{V}{(2\alpha^{\prime})^3}\ ,
\nonumber \\
&&{\rm Re}(T)=\frac{6^{1/3}}{32\pi^3}\frac{V^{1/3}}{2\alpha^{\prime}}\ ,
\eea
where $\phi$ and $\sqrt{2\alpha^{\prime}}$ denote the heterotic
string dilaton and length scale respectively.
One may then expand the effective action of the heterotic string
theory  in powers of the string
loop expansion parameter $\epsilon_s$ and the world--sheet
sigma model expansion parameter $\epsilon_{\sigma}$:   
\bea
&&\epsilon_s = \frac{e^{2\phi}}{(2\pi)^5}
\approx 0.3 \frac{[4\pi^2 {\rm Re}(T)]^3}{{\rm Re}(S)}\ , 
\nonumber \\
&&\epsilon_{\sigma} = \frac{4\pi\alpha^{\prime}}{V^{1/3}}
\approx 0.5 \frac{1}{4\pi^2{\rm Re}(T)}\ .
\label{epsilon}
\eea

Here we are interested in the possible expansion in the $M$-theory limit of 
the strong heterotic string coupling $\epsilon_s\gg 1$ 
for which  $\pi\rho\gtrsim\kappa^{2/9}$ and $V\gtrsim\kappa^{4/3}$ and  so
the physics can be described by 11-dimensional supergravity. 
Since we have two independent length scales, $\rho$ and $V^{1/6}$, 
there can be two dimensionless expansion
parameters in the $M$-theory limit also. 
The analysis of the 11-dimensional theory suggests that
the expansion parameters should scale as $\kappa^{2/3}$ which may be
identified as the inverse of the membrane tension.
One obvious candidate for the expansion parameter in the
$M$-theory limit is the straightforward
generalization of the string world--sheet 
coupling $\sim \alpha^{\prime}/V^{1/3}$ to the membrane
world--volume coupling $\sim \kappa^{2/3}/\rho V^{1/3}$.
Note that in the $M$-theory limit, heterotic string corresponds to 
a membrane stretched along the eleventh dimension.
Information for the other expansion parameter comes from the Witten's
strong coupling expansion of the compactification solution \cite{Witten},
implying that one needs an expansion parameter which is linear in $\rho$.
The analysis of \cite{Witten} shows that
a $\rho$-independent modification of the bulk physics at higher order
in $\kappa^{2/3}$ can lead to a modification of the boundary
physics which is  proportional to $\rho$.
As was noticed in \cite{Banks-Dine,Lukas-Ovrut-Waldram}, this leads to 
an expansion parameter which  scales as $\kappa^{2/3}\rho/V^{2/3}$.
The above observations lead us to  expand 
the 4-dimensional effective SUGRA action of the Ho\v{r}ava-Witten
$M$-theory in powers of
\bea
&&\epsilon_1 = \kappa^{2/3}\pi\rho /V^{2/3}\approx 
 \frac{{\rm Re}(T)}{{\rm Re}(S)}\ , 
\nonumber \\
&&\epsilon_2=\kappa^{2/3}/\pi\rho V^{1/3}\approx 
\frac{1}{4\pi^2{\rm Re}(T)}\ ,
\label{Mexpansion}
\eea
where (1) has been  used  to arrive at this expression of
$\epsilon_1$ and $\epsilon_2$. 
Note that $\epsilon_1\epsilon_2\approx 1/[4\pi^2{\rm Re}(S)]\approx
\alpha_{GUT}/\pi$ which is essentially the 4-dimensional field
theory expansion parameter.
Thus  if one goes to the limit in which one
expansion works better while keeping the realistic value
of $\alpha_{GUT}$,
the other expansion becomes worse.
Here we will simply assume that both $\epsilon_1$ and $\epsilon_2$
are small enough so that  the double expansion in $\epsilon_1$
and $\epsilon_2$  provides
a good perturbative scheme for the effective action
of $M$-theory. 
As we will argue later, this expansion
works well even when $\epsilon_1$ becomes of order one,
which is in fact the case when $M_{GUT}\approx 3\times
10^{16}$ GeV.

Possible guidelines for the $M$-theory expansion of the 4--dimensional
effective action would be
(i) microscopically, the expansion parameter scales as
$\kappa^{2/3}(\pi\rho)^{-n}V^{(n-3)/6}$
and it has a sensible limiting behavior when the Calabi-Yau volume or 
the 11-th segment becomes large, 
(ii) macroscopically, the expansion parameter scales as integral powers
of ${\rm Re}(S)$ and ${\rm Re}(T)$.
In $M$-theory, $V$ can be arbitrarily large
independently of the value of $\pi\rho$.
However as was noted in \cite{Witten}, for a given value of the averaged
Calabi-Yau volume $V$, in order to avoid that one of the boundary
Calabi-Yau volume shrinks to zero, 
$\pi\rho$ is restricted as $\pi\rho\le\kappa^{-2/3}V^{2/3}$.
Then one can demand that the expansion parameter
$\kappa^{2/3}(\pi\rho)^{-n} V^{(n-3)/6}$ should not blow up
in the limit that $V\rightarrow\infty$ for a fixed value of $\pi\rho$
and also in another limit that $\pi\rho\rightarrow\infty$ and
$V\rightarrow\infty$ while keeping $V^{1/6}\ll\pi\rho\le\kappa^{-2/3}V^{2/3}$.
Obviously this condition is satisfied only for $-1\le n\le3$.
There are then only three possible expansion parameters which meet
the guidelines (i) and (ii),
$\epsilon_1$ and $\epsilon_2$ in (\ref{Mexpansion}) and finally
$\epsilon_3=\kappa^{2/3}/(\pi\rho)^3\approx10^{-3}{\rm Re}(S)/[{\rm Re}(T)]^3$
which scales as the inverse of the string loop expansion parameter
$\epsilon_s$.
As will be discussed later, the 4--dimensional effective action computed
in the heterotic string limit can be smoothly extrapolated to
the $M$-theory limit, suggesting that there can be a complete matching
between the heterotic string effective action
(expanded in $\epsilon_s$ and $\epsilon_{\sigma}$)
and the $M$-theory effective action
(expanded in $\epsilon_1$ and $\epsilon_2$).
In view of this matching, it is not likely that there is an additional 
correction which would require to introduce the third expansion parameter
$\epsilon_3$, although we can not rule out this possibility at this moment.
Even when there is such correction,
$\epsilon_3$ is smaller than $\epsilon_2$ and $\epsilon_1$
by one or two orders of magnitude 
for the phenomenologically interesting case
that both ${\rm Re}(S)$ and ${\rm Re}(T)$ are essentially of order one,
which is necessary to have $M_{GUT}\approx 3\times 10^{16}$ GeV together with
the correct value of $\alpha_{GUT}$ (see the discussion below (\ref{gut}).).
This would  justify our scheme of including only the expansions in
$\epsilon_1$ and $\epsilon_2$.

To be explicit, let us consider a simple compactification on  
a Calabi-Yau manifold with the Hodge-Betti number
$h_{1,1}=1$ and $h_{1,2}=0$.
In this model, the low--energy degrees of freedom include first
the gravity multiplet and $S$ and $T$ which are 
the massless modes of
the 11-dimensional bulk fields.
If the spin connection is
embedded into the gauge connection in the observable sector boundary,
we also have the $E_6$ gauge multiplet
and the charged matter multiplet $C$  together with the hidden $E_8$ gauge
multiplet as the massless modes
of the 10-dimensional boundary fields.
It is then easy to compute the K\"ahler potential $K$, the
observable and hidden sector gauge kinetic functions
$f_{E_6}$ and $f_{E_8}$, and the superpotential $W$
at the leading order in the $M$-theory expansion. 
Obviously the leading contribution to the
moduli K\"ahler metric is from 
the  11-dimensional bulk field  action 
which is of order $\kappa^{-2}$, 
while the
charged matter K\"ahler metric,  the gauge kinetic functions, and the
charged matter superpotential receive the leading contributions
from the 10-dimensional boundary action which is
of order $\kappa^{-4/3}$.
Including only the  non-vanishing  leading contributions, one finds
\cite{Banks-Dine,Li-Lopez-Nanopoulos1,Dudas-Grojean,Nilles-Olechowski-Yamaguchi}
\bea
&& K=
-\ln (S+\bar{S})-3\ln (T+\bar{T})+\frac{3|C|^2}{(T+\bar{T})}\ ,
\nonumber \\
&& f_{E_6}=f_{E_8}=S\ , 
\nonumber \\
&& W=d_{pqr}C^pC^qC^r\ ,
\label{tree}
\eea
where $d_{pqr}$ is a constant $E_6$-tensor coefficient.

The holomorphy and the Peccei-Quinn symmetries of (\ref{PQ-symmetry}) 
imply that there is no correction to  the superpotential
at any finite order in the $S$ and $T$-dependent
expansion parameters $\epsilon_1$ and $\epsilon_2$.
However the gauge kinetic functions can receive a correction at order
$\epsilon_1$ in a way  consistent with the holomorphy and the Peccei-Quinn
symmetries.
An explicit computation leads to \cite{Banks-Dine,Choi,Nilles-Stieberger}
\be
f_{E_6}=S+\alpha T\ , \quad f_{E_8}=S-\alpha T\ , 
\label{gaugef}
\ee
where the integer coefficient $\alpha={1\over 8\pi^2}\int\omega\wedge[{\rm tr}
(F\wedge F)-\frac{1}{2}{\rm tr}(R\wedge R)]$
for the K\"ahler form $\omega$ normalized as
the generator of the integer (1,1) 
cohomology\footnote{
Usually $\alpha$ is considered to be an arbitrary real number.
For $T$ normalized as (\protect\ref{dilaton-modulus}),
it is required to be an integer \cite{Choi}.}.
As was discussed in \cite{Banks-Dine,Choi},
the coefficient $\alpha$ in the $M$-theory gauge kinetic function
can be determined either by a direct $M$-theory computation \cite{Witten}
or  by matching the string loop threshold correction
to the gauge kinetic function \cite{Choi-Kim,Ibanez-Nilles}.
At any rate, the holomorphy and the Peccei-Quinn symmetries 
imply that there is no further correction
to the gauge kinetic functions at any finite order
in the $M$-theory expansion.

Let us now consider the possible higher order corrections to
the K\"ahler potential.
With the Peccei-Quinn symmetries, the K\"ahler potential can be written as 
\be
K=\hat{K}(S+\bar{S},T+\bar{T})+Z(S+\bar{S},T+\bar{T})|C|^2\ ,
\ee 
with
\be
\hat{K}=\hat{K}_0+\delta \hat{K}\ ,
\quad
Z=Z_0+\delta Z\ ,
\ee
where $\hat{K}_0=-\ln (S+\bar{S})-3\ln (T+\bar{T})$ and $Z_0=3/(T+\bar{T})$ 
denote the leading order 
results in (\ref{tree}), while $\delta\hat{K}$ and $\delta Z$
are the higher order corrections.
Before going to the $M$-theory expansion of 
$\delta\hat{K}$ and $\delta Z$, it is useful to note that
the bulk physics become blind to the existence of boundaries
in the limit $\rho\rightarrow \infty$.
(This is true up to the trivial scaling of the 4--dimensional
Planck scale $M_P^2\sim \rho$.)
However some of the boundary physics, e.g. the boundary Calabi-Yau
volume, can be affected by the integral of the bulk variables over
the 11-th dimension and then they can include a piece linear in 
$\rho$ \cite{Witten}.
This implies that $\delta\hat{K}/\hat{K}_0$, being the
correction to the pure bulk dynamics, contains only a non-negative
power of $1/\rho$ in the $M$-theory expansion,
while $\delta Z/ Z_0$ which concerns the couplings
between the bulk and boundary fields can include a piece  linear in $\rho$.
Since $\epsilon_1^n\epsilon_2^m\sim \rho^{n-m}$, one needs $m\geq n$
for the expansion of $\delta \hat{K}/\hat{K}_0$ and $m\geq n-1$
for the expansion of $\delta Z/Z_0$.
Taking account of these, the $M$-theory expansion
of the K\"ahler potential  is given by
\bea
\delta \hat{K}&=&\sum_{(n+m\geq 1, m\geq n)}   A_{nm}
\epsilon_1^n\epsilon_2^m \nonumber \\
&=&
\sum_{m\geq 1}\frac{A_{0m}}{[4\pi^2 {\rm Re}(T)]^m}
+\frac{A_{11}}{4\pi^2 {\rm Re}(S)}\left[ 1+{\cal O}(
\frac{1}{4\pi^2{\rm Re}(S)},\frac{1}{4\pi^2{\rm Re}(T)})\right]\ ,
\nonumber \\
\delta Z &= &\frac{3}{(T+\bar{T})}\sum_{(n+m\geq 1, m\geq n-1)}  B_{nm}
\epsilon_1^n\epsilon_2^m \nonumber \\
&=& \frac{3}{(T+\bar{T})}
\sum_{m\geq 1} \frac{B_{0m}}{[4\pi^2 {\rm Re}(T)]^m}
+\frac{3B_{10}}{2{\rm Re}(S)}
\left[1+ {\cal O}(\frac{1}{4\pi^2
{\rm Re}(S)}, \frac{1}{4\pi^2 {\rm Re}(T)})\right]\ ,
\label{expansion}
\eea
where the $n=0$ terms are separated from the other
terms with $n\geq 1$.
Here the coefficients $A_{nm}$ and $B_{nm}$ 
are presumed to be
of order one, and then they 
can have a logarithmic dependence on ${\rm Re}(S)$
or ${\rm Re}(T)$, while all the power-law dependence on ${\rm Re}(S)$
and ${\rm Re}(T)$ 
appear through the expansion parameters $\epsilon_1$ and $\epsilon_2$
which are presumed to be small enough.

The above expansion would work  well 
in the $M$-theory limit: 
\be
[4\pi^2 {\rm Re}(T)]^3\gg {\rm Re}(S) \gg {\rm Re}(T) \gg \frac{1}{4\pi^2}\ ,
\ee
while the heterotic string loop and sigma model
expansions work well in the heterotic string limit: 
\be
{\rm Re}(S)\gg [4\pi^2 {\rm Re}(T)]^3, \quad {\rm Re}(T)\gg \frac{1}{4\pi^2}\ .
\ee
By varying ${\rm Re}(S)$ while keeping ${\rm Re}(T)$
fixed, one can smoothly move 
from the $M$-theory limit 
to the heterotic string limit  (or vice versa)
while keeping $\epsilon_1\approx {\rm Re}(T)/{\rm Re}(S)$ 
and $\epsilon_2\approx 1/[4\pi^2 {\rm Re}(T)]$  small enough.
Obviously then the $M$-theory K\"ahler potential expanded in
$\epsilon_1$ and $\epsilon_2$ remains to be valid over this
procedure, and thus is a  valid expression of the K\"ahler potential
even in the  heterotic
string limit.
This means that, like the case of the gauge kinetic functions,
one can determine the expansion coefficients 
in (\ref{expansion})
by matching  
the heterotic string K\"ahler potential which 
can be computed 
in the string loop and sigma model perturbation theory.
Since $\epsilon_1^n\epsilon_2^m\sim \epsilon_s^n
\epsilon_{\sigma}^{m+2n}$,   $(n,m)$-th order in the $M$-theory expansion
corresponds to $(n,m+2n)$-th order in the string loop and
sigma-model perturbation theory. 
Thus all the terms in the $M$-theory expansion
have their counterparts in the heterotic string expansion.
It appears that  the converse is not true in general,
for instance the term $\epsilon_s^p\epsilon_{\sigma}^q$ with $q<2p$
in the heterotic string expansion 
does  not have its counterpart in the $M$-theory expansion.
However all  string one-loop corrections
which have been computed so far
\cite{Derendinger-Ferrara-Kounnas,Lust,Gava}
lead to corrections which scale
(relative to the leading terms) as  
$\epsilon_s\epsilon_{\sigma}^2$
or $\epsilon_s\epsilon_{\sigma}^3$, 
and thus have $M$-theory counterparts.
This leads us to suspect that all the terms that actually
appear in the heterotic
string expansion have $q\geq 2p$ and thus
have their counterparts in the $M$-theory  
expansion. Then there will be  a complete matching (up
to nonperturbative corrections) of the K\"ahler
potential between the $M$-theory limit and the heterotic string
limit,  like  the case of the gauge kinetic function
and superpotential.

Let us now collect available informations  on the coefficients
in (\ref{expansion}),  either from the heterotic string analysis
or from the direct $M$-theory analysis.   
Clearly in the heterotic string limit, $n$ corresponds
to the number of involved string loops. 
It has been argued that for (2,2) Calabi-Yau compactifications,
a non-vanishing  correction to the K\"ahler potential
starts to occur at 
3-rd order in sigma-model perturbation theory\footnote{ 
For (2,0) compactifications, there may be a correction at lower order,
and then our subsequent discussion should be accordingly modified.}
and thus
\be
A_{01}=A_{02}=B_{01}=B_{02}=0\ ,
\ee
while $A_{03}$ and $B_{03}$ are generically nonvanishing \cite{Choi1}.
The coefficients  $A_{03}$ and $B_{03}$ were explicitly computed
for some (2,2) Calabi-Yau compactifications, yielding for instance 
\cite{Klemm-Theisen,Font,Kim-Munoz}
\begin{equation}
A_{03}=\frac{3}{11}B_{03}= -15\zeta(3)/8\ ,\quad -51\zeta(3)/16\ ,\quad
-111\zeta(3)/16\ ,\quad -27\zeta(3)/2\ ,
\end{equation}
where $\zeta(3)\approx1.2$.
Also the orbifold computation of the string one-loop correction to $\hat{K}$
\cite{Derendinger-Ferrara-Kounnas} suggests that $A_{11}$ is generically
nonvanishing with a possible dependence on $\ln (T+\bar{T})$,
more explicitly $A_{11}\approx \frac{1}{4}\delta_{GS}\ln
(T+\bar{T})$ for the orbifold case
where the Green--Schwarz coefficient $\delta_{GS}$ is generically
of order one.
The  coefficient $B_{10}$ has been recently computed in the context of the
$M$-theory expansion \cite{Lukas-Ovrut-Waldram}, yielding 
\begin{equation}
B_{10}=\frac23\alpha\ .
\end{equation}
Putting these informations altogether, we finally obtain
the following higher order corrections 
to the leading
order K\"ahler potential in (\ref{tree}):
\bea
\delta \hat{K} = && \frac{A_{03}}{[4\pi^2 {\rm Re}(T)]^3}
\left[ 1+{\cal O}(
\frac{1}{4\pi^2{\rm Re}(T)})\right] \nonumber
\\
&&+\frac{A_{11}}{4\pi^2 {\rm Re}(S)}
\left[ 1+{\cal O}(
\frac{1}{4\pi^2{\rm Re}(S)},\frac{1}{4\pi^2{\rm Re}(T)})\right],
\nonumber \\
\delta Z = && 
\frac{3}{(T+\bar{T})}\frac{B_{03}}{[4\pi^2 {\rm Re}(T)]^3}
\left[ 1+{\cal O}(
\frac{1}{4\pi^2{\rm Re}(T)})\right] \nonumber \\
&&+\frac{\alpha}{{\rm Re}(S)}
\left[ 1+{\cal O}(
\frac{1}{4\pi^2{\rm Re}(S)},\frac{1}{4\pi^2{\rm Re}(T)})\right].
\label{correction}
\eea 

As a phenomenological application of the $M$-theory expansion
discussed so far, in the following we are going to
analyze the soft SUSY--breaking terms 
under the assumption that SUSY is spontaneously broken by
the auxiliary components $F^S$ and $F^T$ of the moduli superfields
$S$ and $T$.  
To simplify the analysis, we will  concentrate on the
moduli values leading to  $M_{GUT}\approx 3\times 10^{16}$ GeV.
Using 
(\ref{dilaton-modulus}) and
the $M$-theory expression of the 4-dimensional Planck scale,
$M_P^2=2\pi\rho V/\kappa^2$, one easily finds
\be
M_{GUT}=1/V^{1/6}\approx 3.6\times 
10^{16} \left(\frac{2}{{\rm Re}(S)}\right)^{1/2}
\left(\frac{1}{{\rm Re}(T)}\right)^{1/2} \, {\rm GeV}\ .
\label{gut}
\ee
Since from (\ref{gaugef}) one obtains
${\rm Re}(S)+\alpha {\rm Re}(T)={\rm Re}f_{E_6}=1/g_{GUT}^2\approx 2$,  
the above relation 
implies  that ${\rm Re}(T)$ is essentially
of order one  
when  $M_{GUT}\approx 3\times 10^{16}$ GeV. 
Clearly, if ${\rm Re}(T)$ is of order one,
we are in the $M$-theory domain with  $\epsilon_s\gg 1$. (See 
(\ref{epsilon})).
One may  
worry  that the $M$-theory expansion (\ref{expansion}) would  not work
in this case
since 
$\epsilon_1={\rm Re}(T)/{\rm Re}(S)$ is of order one also.
However as we have noticed, any correction which is $n$-th order
in $\epsilon_1$ accompanies at least $(n-1)$-powers of
$\epsilon_2$ and thus is suppressed by $(\epsilon_1\epsilon_2)^{n-1}
\approx (\alpha_{GUT}/\pi)^{n-1}$ compared to
the order $\epsilon_1$ correction.
This allows the $M$-theory expansion (\ref{expansion}) to be valid even when
$\epsilon_1$ becomes of order one.
Obviously if ${\rm Re}(T)$ is of order one, 
only the order $\epsilon_1$ correction to $Z$,
i.e. $\delta Z=\alpha/{\rm Re}(S)$, can be sizable.
The other corrections are  suppressed  by  either 
$\epsilon_1\epsilon_2\approx 1/4\pi^2{\rm Re}(S)$ or $\epsilon_2^3
\approx 1/[4\pi^2 {\rm Re}(T)]^3$ and thus smaller than
the leading order results at least by  ${\cal O}(\frac{\alpha_{GUT}}{\pi})$.
Thus we will include  only $\delta Z=\alpha/{\rm Re}(S)$ in the later
analysis of soft terms, while ignoring
the other corrections to the K\"ahler potential\footnote{
In fact, in the weakly coupled heterotic string limit, it is quite
possible that $1/4\pi^2{\rm Re}(T)$ is {\it not} so small,
and then the sigma model corrections of order $1/[4\pi^2 {\rm Re}(T)]^3$
can significantly affect the soft terms \cite{Choi1,Kim-Munoz}.
However in the $M$-theory limit,  $1/4\pi^2{\rm Re}(T)$ is quite small
and thus  the effects of these sigma model corrections are negligible
compared to those of $\delta Z=\alpha/{\rm Re}(S)$.}.

Summarizing the above discussion, our starting point of the soft term analysis
is the effective SUGRA model given by
\bea
&&K=-\ln (S+\bar{S}) -3\ln (T+\bar{T})+\left(\frac{3}{T+\bar{T}}+
\frac{\alpha}{S+\bar{S}}\right) |C|^2\ ,
\nonumber \\
&&f_{E_6}=S+\alpha T\ , \quad f_{E_8}=S-\alpha T\ , 
\nonumber \\
&&W=d_{pqr}C^pC^qC^r\ .
\label{effective-SUGRA}
\eea
Here the superpotential and gauge kinetic functions are exact up to 
nonperturbative corrections, while there can be small additional  perturbative
corrections to the K\"ahler potential
which are of  order $1/4\pi^2 {\rm Re}(S)$ or $1/[4\pi^2 {\rm Re}(T)]^3$.
Since the above SUGRA model includes only a single $T$-modulus, 
the resulting soft terms would  describe 
the case that  only one combination of the  $T$-moduli, if there are 
more than one,  participates in the SUSY breaking.
Later, we will also discuss the multimoduli case.

Applying  the standard (tree--level) 
soft term formulae \cite{Soni-Weldon,Brignole-Ibanez-Munoz}
for the above SUGRA model (\ref{effective-SUGRA}), 
we can compute the soft terms straightforwardly. 
Since the bilinear parameter, $B$, depends on the specific mechanism which
could generate the associated $\mu$ term, let us concentrate on
gaugino masses, $M$, scalar masses, $m$, and trilinear parameters, $A$.
After normalizing the observable fields, these are
given by
\bea
M =&& \frac{\sqrt 3 Cm_{3/2}}
{(S+\bar S) + \alpha (T+ \bar T)}
\left((S+\bar S)\sin\theta e^{-i\gamma_S}+ \frac{\alpha(T+\bar T)}{\sqrt 3}
\cos\theta e^{-i\gamma_T}\right)
\ , \nonumber \\
m^2 =&& V_0 + m_{3/2}^2 - \frac{3m_{3/2}^2 C^2}{ 
3(S+\bar S)+\alpha (T+\bar T)}
\nonumber\\ &&
\times \left\{
\alpha (T+\bar T) \left(2-\frac{\alpha (T+\bar T)}
{3(S+\bar S) + \alpha (T+\bar T)}\right) \sin^2\theta
\right.  \nonumber\\ &&
+ (S+\bar S) \left(2-\frac{3(S+\bar S)}
{3(S+\bar S) + \alpha (T+\bar T)}\right) \cos^2\theta
\nonumber\\ && - \left.
\frac{2\sqrt 3 \alpha (T+\bar T)(S+ \bar S)}
{3(S+\bar S)+ \alpha (T+\bar T)}
\sin\theta \cos\theta \cos(\gamma_S-\gamma_T)
\right\} \ , \nonumber\\
A =&& \sqrt 3 C m_{3/2} \left\{
\left(-1 + \frac{3\alpha (T+ \bar T)}
{3(S+\bar S)+\alpha (T+\bar T)}\right) 
\sin\theta e^{-i\gamma_S}\right.
\nonumber\\ &&
+ \left.  \sqrt 3 \left(-1 + \frac{3(S+\bar S)}
{3(S+\bar S) + \alpha (T+\bar T)}\right)
\cos\theta e^{-i\gamma_T} \right\}\ ,
\label{ssoft}
\eea
where we are using the parametrization 
introduced in \cite{Brignole-Ibanez-Munoz2}
in order to know what fields, either $S$ or $T$, play the predominant role
in the process of SUSY breaking
\bea
F^S &=& \sqrt 3 m_{3/2}C(S+\bar S)\sin\theta e^{-i\gamma_S}\ ,
\nonumber \\
F^T &=& m_{3/2}C(T+\bar T)\cos\theta e^{-i\gamma_T}\ ,
\label{fterms}
\eea
with $m_{3/2}$ for the gravitino mass, $C^2=1+V_0/3m_{3/2}^2$ 
and $V_0$ for the tree-level vacuum energy density.
In what follows, given the current experimental limits, 
we will assume $V_0=0$
and $\gamma_S=\gamma_T=0\ ({\rm mod}\ \, \pi)$. 
More specifically, we will set $\gamma_S$ and $\gamma_T$ 
to zero and allow $\theta$ to vary in a range $[0, 2\pi)$. 

Notice that the structure of these soft terms is qualitatively different 
from those of the weakly coupled heterotic string case
\cite{Brignole-Ibanez-Munoz2} which can be recovered
from (\ref{ssoft}) by 
taking the limit\footnote{
Of course, in this limit the corrections of
order $1/[4\pi^2{\rm Re}(T)]^3$ in (\ref{correction}) which are ignored
in (\ref{effective-SUGRA}) can be important, however we will
ignore  this point for simplicity.}
$\alpha(T+\bar T)\ll (S+\bar S)$:
%
%
\bea
M = -A = \sqrt 3 m_{3/2} \sin\theta\ , \quad 
m^2=m_{3/2}^2\left(1-\cos^2\theta\right)\ .
\label{wsoft}
\eea
Whereas 
there are only two free parameters in (\ref{wsoft}), viz $m_{3/2}$ and
$\theta$, the $M$--theory result (\ref{ssoft}) is more involved 
due to the additional dependence on ($S+\bar S$) and 
$\alpha(T+\bar T)$ even when we set $C=1$ and $\gamma_S=\gamma_T=0$.
As a consequence, even in the dilaton--dominated SUSY--breaking scenario
with $|\sin\theta|=1$, it is no longer true that  simple results
$M=-A=\pm \sqrt 3 m$ and $m=m_{3/2}$ hold. 
(We will discuss in more detail this scenario below.) 
Nevertheless we can simplify the analysis by taking into account,
as already mentioned above,
that the real part 
of the gauge kinetic functions in (\ref{effective-SUGRA}) 
are the inverse squared gauge coupling constants and thus
\bea
(S+\bar S)
+\alpha(T+ \bar T) 
\approx 4
\label{bound1}
\eea
to produce the known values of $\alpha_{GUT}$.
We then have only one more parameter, say $\alpha(T+\bar T)$,
than the result (\ref{wsoft}) in the heterotic string limit.
In fact, this parameter is severely constrained
by ${\rm Re}(f_{E_8})>0$, leading to $\alpha (T+\bar T) < (S+\bar S)$.
Using (\ref{bound1}), we then  obtain the following bound 
\bea
0< \alpha(T+\bar T) \lesssim 2\ . 
\label{bound2}
\eea

Given these results, and recalling the analysis of the GUT scale
(\ref{gut}) where we obtained that ${\rm Re}(T)$ should be of order one, 
we show in Fig.~1 the dependence on $\theta$ of
the soft terms $M$, $m$, and $A$ in units of the gravitino mass for
different values of $\alpha(T+\bar T)$. 
Several comments are in order.
First of all, some ranges of $\theta$ are forbidden
by having a  negative scalar mass-squared.
The figures clearly show that the smaller the value of  $\alpha(T+\bar T)$,  
the smaller the forbidden regions become.
In the weakly coupled heterotic string case
shown in Fig.~2, the forbidden region vanishes since
the squared scalar masses are always positive (see (\ref{wsoft})).
Notice however that even in the extreme case
$\alpha(T+\bar T)=2$, shown in Fig.~1(a), 
the allowed regions correspond to values
of $\theta$ such that
$|\sin\theta|<0.9$ and therefore 
most of the dilaton/modulus SUSY--breaking scenarios are possible.
About the possible range of soft terms,
the smaller the value of $\alpha(T+\bar T)$, the larger the range becomes.
For example, for $\alpha(T+\bar{T})=2$ (Fig.~1(a)), those ranges are
$0.5<|M|/m_{3/2}<1$, $0<m/m_{3/2}<0.5$ and $0.71<|A|/m_{3/2}<0.87$,
whereas for 
$\alpha(T+\bar T)=1$ (Fig.~1(c)), they are  
$0.25<|M|/m_{3/2}<1.32$, $0<m/m_{3/2}<0.7$ and $0.3<|A|/m_{3/2}<1.25$.

In order to discuss the SUSY spectra further, it is worth
noticing  that scalar masses are always smaller than gaugino masses. 
This is shown in Fig.~3 where the ratio $m/|M|$ versus $\theta$ is plotted 
for different values of $\alpha(T+\bar T)$.
Notice that in the heterotic string limit
which corresponds to the straight line $m/|M|=1/\sqrt 3$, 
the limit $\sin\theta\rightarrow 0$ is not well defined 
since all $M$, $A$, $m$  vanish in that limit. 
One then has to include  the string one--loop 
corrections (or the sigma--model corrections) 
to the K\"ahler potential and gauge kinetic
functions which would modify the boundary conditions (\ref{wsoft}).
This problem is not present in the $M$--theory
limit since  gaugino masses are always different from zero.

Finally, given the numerical results and also Fig.~1 and Fig.~2,
it is straightforward to compare
the $M$--theory limit with the weakly coupled heterotic string 
limit  in the dilaton--dominated case with 
$|\sin\theta|=1$.
For example, whereas $m^2=m_{3/2}^2$ in the heterotic string limit,
now in the $M$-theory limit, $m^2$  can be much smaller and even
negative. See Fig.~1(a) with respect to the possibility 
of a negative $m^2$ in the dilaton--dominated scenario,
where  $|\sin\theta|=1$ is excluded.
For a further comparison, let us consider the case when several moduli $T_i$
and the associated matter $C_i$ are present \cite{Kim-Munoz}.
The soft scalar  masses in this case are given by
\begin{eqnarray}
{m}^2_{{i}{\bar{j}}} &=&
m_{3/2}^2 Z_{i\bar j}
- {F}^{m}\left( 
\partial_m\partial_{\bar{n}}
{Z_{{i}{\bar{j}}}}
-{Z}^{k\bar{l}}\partial_m {Z}_{i
\bar{l}}
\partial_{\bar{n}}{Z}_{k\bar{j}}  
\right){\bar F}^{\bar n} \ ,
\label{universal}
\end{eqnarray}
where $F^m=F^S, F^{T_i}$, and $Z_{i\bar{j}}$ and
$Z^{i\bar{j}}$ denote the K\"ahler metric and its inverse
of the matter fields $C_i$. 
After normalizing the fields to get canonical kinetic terms, although the
first piece in (\ref{universal}) will lead to universal diagonal soft masses,
the second piece will generically induce non--universal contributions 
due to the presence of off--diagonal K\"ahler metric:
\begin{eqnarray}
Z_{i\bar j} = 
({\partial}^2 {\hat K}^T/\partial T_i \partial {\bar T}_j)
e^{-{\hat K}^T/3}+\delta Z_{i\bar j}(S+\bar{S},T_i+\bar{T}_i)\ ,
\end{eqnarray}
where 
\begin{eqnarray}
{\hat K}^T=
-\ln k_{ijk} 
(T_i+{\bar T}_i)
(T_j+{\bar T}_j)
(T_k+{\bar T}_k)
\end{eqnarray}
and $\delta Z_{i\bar j}$ corresponds to
the $S$-dependent correction in the $M$-theory expansion (or
the string-loop correction).
If one ignores $\delta Z_{i\bar j}$, the matter K\"ahler metric
is $S$-independent, and as a consequence
in the dilaton--dominated scenario with
$F^{T_i}=0$, 
the normalized soft scalar masses are   
universal as $m_i=m_{3/2}$. 
However including the $S$-dependent $\delta Z_{i\bar j}$,
one generically loses the scalar mass universality 
even in the dilaton--dominated case.
In fact, this 
was noted   in \cite{Nir} 
for the string--loop induced $\delta Z_{i\bar j}$
which is small  in the weakly coupled heterotic string limit.
Our main point here
is that in the $M$-theory limit $\delta Z_{i\bar j}$
can be as large as the 
leading order K\"ahler metric,
and then 
there can be a large violation
of the scalar mass universality even in the dilaton--dominated 
scenario.

It is worth noticing here that the above comments about the dilaton--dominated
scenario may also be applied  to
orbifold compactifications assuming that similar $M$--theory effects
are present.
For example,
the K\"ahler metric of untwisted matter field $C_i$ 
may be given by
\be
Z_{i} = \frac{1}{T_i+{\bar T}_i}
+ \frac{\alpha_i}{S+\bar S}
\ ,
\ee
where the second piece corresponds to the high order correction
in the $M$-theory expansion.
Then the soft scalar mass of $C_i$ in the dilaton--dominated scenario 
can be obtained
from (\ref{ssoft})  with the substitution 
$\alpha(T+\bar T)\rightarrow 3\alpha_i(T_i+\bar T_i)$ and
$|\sin\theta|=1$, which shows clearly that
the scalar mass
universality is lost. This is also true for 
$\alpha_i=\alpha$ since still the VEVs of the $T_i$'s will be different
in general.

Let us now discuss the 
predictions for the low--energy ($\approx M_W$) sparticle
spectra in this 
$M$--theory scenario. As is well known 
there are several particles whose masses are rather independent of the
details of $SU(2)_L\times U(1)_Y$ radiative breaking and are mostly given
by the boundary conditions and the renormalization group running
\cite{Brignole-Ibanez-Munoz}.
In particular, that is the case of the gluino $\tilde g$, 
all the sleptons $\tilde l$ and
first and second generation squarks $\tilde q$.
Since, as discussed above, always 
$m<|M|$ at the GUT scale, 
the qualitative mass relations at the electroweak scale in  the $M$-theory
scenario turn out to be
\bea
M_{\tilde g}\approx m_{\tilde q}>m_{\tilde l}\ , 
\label{mass}
\eea
where gluinos are 
slightly heavier than squarks. We recall that slepton masses are smaller
than squark masses because they do not feel the important gluino
contribution in the renormalization. The precise values
in (\ref{mass}) depend on the ratio $r\equiv m/|M|$:
\bea
&& M_{\tilde g}:m_{\tilde Q_L}:m_{\tilde u_R}:m_{\tilde d_R}:m_{\tilde L_L}:
m_{\tilde e_R} \nonumber \\
&&\approx 
1:\frac{1}{3}\sqrt {7.6+r^2}:\frac{1}{3}\sqrt {7.17+r^2}:
\frac{1}{3}\sqrt {7.14+r^2}:\frac{1}{3}\sqrt{0.53+r^2}:\frac{1}{3}
\sqrt{0.15+r^2}.
\eea
For example for 
$r=1/\sqrt 3$, which is always the case of the weakly coupled
heterotic string  limit, one obtains 
$1:0.94:0.92:0.91:0.3:0.23$, 
whereas for the extreme case
of $r=0$ the result is 
$1:0.92:0.89:0.88:0.24:0.13$.
Clearly, this type of analysis 
would allow us to distinguish the
$M$--theory limit from the weakly coupled heterotic string limit.
If the observed SUSY spectrum is inconsistent with the above results
for $r=1/\sqrt 3$,
the M--theory limit may be the answer. On the other hand, 
we see in Fig.~3 that 
$r=1/\sqrt 3$ can  be obtained for particular values of $\theta$
in the M--theory limit also.
Thus if the SUSY spectrum turns out to be  consistent with $r=1/\sqrt 3$,
one should analyze 
the rest of the SUSY mass spectra, taking into account the
details of the electroweak radiative breaking. This more detailed analysis
would allow us to distinguish clearly between both limits.
To this respect, we note  that even for $r$ which is close to $1/\sqrt 3$,
the pattern of soft terms in the  $M$--theory limit significantly differs
from that in the heterotic string limit (see Figs.~1, 2).
Although most of our analysis has been made for the simple case
that only $S$ and one of the possible $T$-moduli participate
in SUSY breaking,
this kind of analysis can be easily generalized to a more general case.


\bigskip

{\bf Acknowledgments}:
This work is supported in part by
KAIST Basic Science Research Program (KC),
KAIST Center for Theoretical Physics and Chemistry (KC, HBK),
KOSEF through CTP of Seoul National University (KC),
the KRF under the Distinguished Scholar Exchange Program (KC),
Basic Science Research Institutes Program BSRI-97-2434 (KC),
the KOSEF under the Brainpool Program (CM), and
the CICYT under contract AEN97-1678-E (CM).

%
%

\def\softplot#1#2{%
\beginpicture
\setcoordinatesystem units <100pt,40pt> point at 0 -2
\setplotarea x from 0.0 to 2.0, y from -2.0 to 2.0
\inboundscheckon
\linethickness 0.5pt
\axis bottom label {}
      ticks in
      width <0.5pt> length <3.0pt> unlabeled quantity 21
      width <0.5pt> length <6.0pt>
      withvalues {$0$} {$\frac12\pi$} {$\pi$} {$\frac32\pi$} {$2\pi$} /
      at 0 0.5 1.0 1.5 2.0 /
/
\axis left label {}
      ticks in
      width <0.5pt> length <6.0pt>
      at -1.5 -0.5 0.5 1.5 /
      width <0.5pt> length <6.0pt>
      withvalues {-2} {-1} {0} {1} {2} /
      at -2.0 -1.0 0.0 1.0 2.0 /
/
\axis top label {}
      ticks in
      width <0.5pt> length <3.0pt> unlabeled quantity 21
      width <0.5pt> length <6.0pt>
      at 0 0.5 1.0 1.5 2.0 /
/
\axis right label {}
      ticks in
      width <0.5pt> length <6.0pt>
      at -2.0 -1.5 -1.0 -0.5 0.0 0.5 1.0 1.5 2.0 /
/
\setplotsymbol ({\tiny.})
\setlinear
\plot 0.0 0.0 2.0 0.0 /
\multiput {\vrule height 4pt} [c] at 0 0 *20 0.1 0 /
\multiput {\vrule height 8pt} [c] at 0 0 *4 0.5 0 /
#1
\setplotsymbol ({.})
\setquadratic #2
\endpicture
}

\def\plota{\softplot{%
\put {(a)} <18pt,-15pt> at 0.0 2.0
\put {$\alpha(T+\bar T)=2.0$} <0pt,-15pt> at 1.0 2.0
\put {$M$} at 0.43  1.00
\put {$m$} at 0.43  0.20
\put {$A$} at 0.43 -0.65
\put {$M$} at 1.43 -1.00
\put {$m$} at 1.43  0.20
\put {$A$} at 1.43  0.75
}{
\setdots <2pt>
\plot
 0.00000   0.50000 
 0.01000   0.52696 
 0.02000   0.55339 
 0.03000   0.57928 
 0.04000   0.60460 
 0.05000   0.62932 
 0.06000   0.65342 
 0.07000   0.67688 
 0.08000   0.69966 
 0.09000   0.72176 
 0.10000   0.74314 
 0.11000   0.76380 
 0.12000   0.78369 
 0.13000   0.80282 
 0.14000   0.82115 
 0.15000   0.83867 
 0.16000   0.85536 
 0.17000   0.87121 
 0.18000   0.88620 
 0.19000   0.90032 
 0.20000   0.91355 
 0.21000   0.92587 
 0.22000   0.93728 
 0.23000   0.94777 
 0.24000   0.95732 
 0.25000   0.96593 
 0.26000   0.97358 
 0.27000   0.98027 
 0.28000   0.98600 
 0.29000   0.99075 
 0.30000   0.99452 
 0.31000   0.99731 
 0.32000   0.99912 
 0.33000   0.99995 
 0.34000   0.99978 
 0.35000   0.99863 
 0.36000   0.99649 
/
\plot
 0.98000  -0.44464 
 0.99000  -0.47255 
 1.00000  -0.50000 
 1.01000  -0.52696 
 1.02000  -0.55339 
 1.03000  -0.57928 
 1.04000  -0.60460 
 1.05000  -0.62932 
 1.06000  -0.65342 
 1.07000  -0.67688 
 1.08000  -0.69966 
 1.09000  -0.72176 
 1.10000  -0.74314 
 1.11000  -0.76380 
 1.12000  -0.78369 
 1.13000  -0.80282 
 1.14000  -0.82115 
 1.15000  -0.83867 
 1.16000  -0.85536 
 1.17000  -0.87121 
 1.18000  -0.88620 
 1.19000  -0.90032 
 1.20000  -0.91355 
 1.21000  -0.92587 
 1.22000  -0.93728 
 1.23000  -0.94777 
 1.24000  -0.95732 
 1.25000  -0.96593 
 1.26000  -0.97358 
 1.27000  -0.98027 
 1.28000  -0.98600 
 1.29000  -0.99075 
 1.30000  -0.99452 
 1.31000  -0.99731 
 1.32000  -0.99912 
 1.33000  -0.99995 
 1.34000  -0.99978 
 1.35000  -0.99863 
 1.36000  -0.99649 
/
\plot
 1.98000   0.44464 
 1.99000   0.47255 
 2.00000   0.50000 
/
\setsolid
\plot
 0.00000   0.25000 
 0.01000   0.28727 
 0.02000   0.31894 
 0.03000   0.34646 
 0.04000   0.37064 
 0.05000   0.39202 
 0.06000   0.41096 
 0.07000   0.42770 
 0.08000   0.44245 
 0.09000   0.45533 
 0.10000   0.46645 
 0.11000   0.47590 
 0.12000   0.48373 
 0.13000   0.48999 
 0.14000   0.49472 
 0.15000   0.49794 
 0.16000   0.49967 
 0.17000   0.49992 
 0.18000   0.49868 
 0.19000   0.49596 
 0.20000   0.49174 
 0.21000   0.48599 
 0.22000   0.47869 
 0.23000   0.46979 
 0.24000   0.45923 
 0.25000   0.44694 
 0.26000   0.43283 
 0.27000   0.41677 
 0.28000   0.39859 
 0.29000   0.37806 
 0.30000   0.35486 
 0.31000   0.32853 
 0.32000   0.29836 
 0.33000   0.26317 
 0.34000   0.22076 
 0.35000   0.16591 
 0.36000   0.07559 
 0.36100   0.05919
 0.36258   0.00000
/
\plot
 0.97075   0.00000
 0.97100   0.02340
 0.97200   0.05259
 0.97300   0.07056
 0.97400   0.08477
 0.97500   0.09689
 0.97600   0.10763
 0.97700   0.11737
 0.97800   0.12634
 0.97900   0.13470
 0.98000   0.14254 
 0.99000   0.20430 
 1.00000   0.25000 
 1.01000   0.28727 
 1.02000   0.31894 
 1.03000   0.34646 
 1.04000   0.37064 
 1.05000   0.39202 
 1.06000   0.41096 
 1.07000   0.42770 
 1.08000   0.44245 
 1.09000   0.45533 
 1.10000   0.46645 
 1.11000   0.47590 
 1.12000   0.48373 
 1.13000   0.48999 
 1.14000   0.49472 
 1.15000   0.49794 
 1.16000   0.49967 
 1.17000   0.49992 
 1.18000   0.49868 
 1.19000   0.49596 
 1.20000   0.49174 
 1.21000   0.48599 
 1.22000   0.47869 
 1.23000   0.46979 
 1.24000   0.45923 
 1.25000   0.44694 
 1.26000   0.43283 
 1.27000   0.41677 
 1.28000   0.39859 
 1.29000   0.37806 
 1.30000   0.35486 
 1.31000   0.32853 
 1.32000   0.29836 
 1.33000   0.26317 
 1.34000   0.22076 
 1.35000   0.16591 
 1.36000   0.07559 
 1.36100   0.05919
 1.36258   0.00000
/
\plot
 1.97075   0.00000
 1.97100   0.02340
 1.97200   0.05259
 1.97300   0.07056
 1.97400   0.08477
 1.97500   0.09689
 1.97600   0.10763
 1.97700   0.11737
 1.97800   0.12634
 1.97900   0.13470
 1.98000   0.14254 
 1.99000   0.20430 
 2.00000   0.25000 
/
\setdashes <2pt>
\plot
 0.00000  -0.75000 
 0.01000  -0.76323 
 0.02000  -0.77571 
 0.03000  -0.78742 
 0.04000  -0.79836 
 0.05000  -0.80850 
 0.06000  -0.81785 
 0.07000  -0.82640 
 0.08000  -0.83412 
 0.09000  -0.84103 
 0.10000  -0.84710 
 0.11000  -0.85234 
 0.12000  -0.85673 
 0.13000  -0.86029 
 0.14000  -0.86299 
 0.15000  -0.86484 
 0.16000  -0.86584 
 0.17000  -0.86598 
 0.18000  -0.86527 
 0.19000  -0.86370 
 0.20000  -0.86128 
 0.21000  -0.85801 
 0.22000  -0.85390 
 0.23000  -0.84894 
 0.24000  -0.84314 
 0.25000  -0.83652 
 0.26000  -0.82906 
 0.27000  -0.82079 
 0.28000  -0.81171 
 0.29000  -0.80183 
 0.30000  -0.79115 
 0.31000  -0.77970 
 0.32000  -0.76747 
 0.33000  -0.75449 
 0.34000  -0.74077 
 0.35000  -0.72631 
 0.36000  -0.71114 
/
\plot
 0.98000   0.72133 
 0.99000   0.73603 
 1.00000   0.75000 
 1.01000   0.76323 
 1.02000   0.77571 
 1.03000   0.78742 
 1.04000   0.79836 
 1.05000   0.80850 
 1.06000   0.81785 
 1.07000   0.82640 
 1.08000   0.83412 
 1.09000   0.84103 
 1.10000   0.84710 
 1.11000   0.85234 
 1.12000   0.85673 
 1.13000   0.86029 
 1.14000   0.86299 
 1.15000   0.86484 
 1.16000   0.86584 
 1.17000   0.86598 
 1.18000   0.86527 
 1.19000   0.86370 
 1.20000   0.86128 
 1.21000   0.85801 
 1.22000   0.85390 
 1.23000   0.84894 
 1.24000   0.84314 
 1.25000   0.83652 
 1.26000   0.82906 
 1.27000   0.82079 
 1.28000   0.81171 
 1.29000   0.80183 
 1.30000   0.79115 
 1.31000   0.77970 
 1.32000   0.76747 
 1.33000   0.75449 
 1.34000   0.74077 
 1.35000   0.72631 
 1.36000   0.71114 
/
\plot
 1.98000  -0.72133 
 1.99000  -0.73603 
 2.00000  -0.75000 
/
}}

\def\plotb{\softplot{%
\put {(b)} <18pt,-15pt> at 0.0 2.0
\put {$\alpha(T+\bar T)=1.5$} <0pt,-15pt> at 1.0 2.0
\put {$M$} at 0.62  1.00
\put {$m$} at 0.62  0.20
\put {$A$} at 0.62 -0.65
\put {$M$} at 1.62 -1.00
\put {$m$} at 1.62  0.20
\put {$A$} at 1.62  0.75
}{
\setdots <2pt>
\plot
 0.00000   0.37500 
 0.01000   0.40882 
 0.02000   0.44223 
 0.03000   0.47521 
 0.04000   0.50772 
 0.05000   0.53973 
 0.06000   0.57120 
 0.07000   0.60212 
 0.08000   0.63243 
 0.09000   0.66213 
 0.10000   0.69117 
 0.11000   0.71952 
 0.12000   0.74717 
 0.13000   0.77408 
 0.14000   0.80023 
 0.15000   0.82559 
 0.16000   0.85013 
 0.17000   0.87383 
 0.18000   0.89667 
 0.19000   0.91863 
 0.20000   0.93968 
 0.21000   0.95980 
 0.22000   0.97897 
 0.23000   0.99718 
 0.24000   1.01441 
 0.25000   1.03063 
 0.26000   1.04584 
 0.27000   1.06001 
 0.28000   1.07314 
 0.29000   1.08521 
 0.30000   1.09621 
 0.31000   1.10612 
 0.32000   1.11495 
 0.33000   1.12267 
 0.34000   1.12929 
 0.35000   1.13479 
 0.36000   1.13917 
 0.37000   1.14243 
 0.38000   1.14456 
 0.39000   1.14556 
 0.40000   1.14543 
 0.41000   1.14417 
 0.42000   1.14178 
 0.43000   1.13826 
 0.44000   1.13363 
 0.45000   1.12787 
 0.46000   1.12100 
 0.47000   1.11302 
 0.48000   1.10394 
 0.49000   1.09378 
 0.50000   1.08253 
 0.51000   1.07022 
 0.52000   1.05685 
 0.53000   1.04244 
 0.54000   1.02700 
 0.55000   1.01054 
 0.55000   1.01054 
/
\plot
 0.99000  -0.34081 
 1.00000  -0.37500 
 1.01000  -0.40882 
 1.02000  -0.44223 
 1.03000  -0.47521 
 1.04000  -0.50772 
 1.05000  -0.53973 
 1.06000  -0.57120 
 1.07000  -0.60212 
 1.08000  -0.63243 
 1.09000  -0.66213 
 1.10000  -0.69117 
 1.11000  -0.71952 
 1.12000  -0.74717 
 1.13000  -0.77408 
 1.14000  -0.80023 
 1.15000  -0.82559 
 1.16000  -0.85013 
 1.17000  -0.87383 
 1.18000  -0.89667 
 1.19000  -0.91863 
 1.20000  -0.93968 
 1.21000  -0.95980 
 1.22000  -0.97897 
 1.23000  -0.99718 
 1.24000  -1.01441 
 1.25000  -1.03063 
 1.26000  -1.04584 
 1.27000  -1.06001 
 1.28000  -1.07314 
 1.29000  -1.08521 
 1.30000  -1.09621 
 1.31000  -1.10612 
 1.32000  -1.11495 
 1.33000  -1.12267 
 1.34000  -1.12929 
 1.35000  -1.13479 
 1.36000  -1.13917 
 1.37000  -1.14243 
 1.38000  -1.14456 
 1.39000  -1.14556 
 1.40000  -1.14543 
 1.41000  -1.14417 
 1.42000  -1.14178 
 1.43000  -1.13826 
 1.44000  -1.13363 
 1.45000  -1.12787 
 1.46000  -1.12100 
 1.47000  -1.11302 
 1.48000  -1.10394 
 1.49000  -1.09378 
 1.50000  -1.08253 
 1.51000  -1.07022 
 1.52000  -1.05685 
 1.53000  -1.04244 
 1.54000  -1.02700 
 1.55000  -1.01054 
/
\plot
 1.99000   0.34081 
 2.00000   0.37500 
 2.00000   0.37500 
/
\setsolid
\plot
 0.00000   0.16667 
 0.01000   0.20721 
 0.02000   0.24114 
 0.03000   0.27083 
 0.04000   0.29745 
 0.05000   0.32168 
 0.06000   0.34393 
 0.07000   0.36448 
 0.08000   0.38354 
 0.09000   0.40125 
 0.10000   0.41771 
 0.11000   0.43300 
 0.12000   0.44719 
 0.13000   0.46033 
 0.14000   0.47245 
 0.15000   0.48358 
 0.16000   0.49375 
 0.17000   0.50297 
 0.18000   0.51127 
 0.19000   0.51865 
 0.20000   0.52513 
 0.21000   0.53071 
 0.22000   0.53540 
 0.23000   0.53920 
 0.24000   0.54212 
 0.25000   0.54417 
 0.26000   0.54533 
 0.27000   0.54562 
 0.28000   0.54503 
 0.29000   0.54357 
 0.30000   0.54123 
 0.31000   0.53801 
 0.32000   0.53390 
 0.33000   0.52891 
 0.34000   0.52302 
 0.35000   0.51624 
 0.36000   0.50855 
 0.37000   0.49993 
 0.38000   0.49039 
 0.39000   0.47989 
 0.40000   0.46843 
 0.41000   0.45597 
 0.42000   0.44248 
 0.43000   0.42792 
 0.44000   0.41224 
 0.45000   0.39536 
 0.46000   0.37721 
 0.47000   0.35766 
 0.48000   0.33655 
 0.49000   0.31366 
 0.50000   0.28868 
 0.51000   0.26110 
 0.52000   0.23014 
 0.53000   0.19433 
 0.54000   0.15045 
 0.55000   0.08738 
 0.55100   0.07839
 0.55200   0.06824
 0.55300   0.05631
 0.55400   0.04106
 0.55514   0.00000
/
\plot
 0.98146   0.00000
 0.98200   0.02839
 0.98300   0.04785
 0.98400   0.06144
 0.98500   0.07254
 0.98600   0.08217
 0.98700   0.09079
 0.98800   0.09867
 0.98900   0.10598
 0.99000   0.11282 
 1.00000   0.16667 
 1.01000   0.20721 
 1.02000   0.24114 
 1.03000   0.27083 
 1.04000   0.29745 
 1.05000   0.32168 
 1.06000   0.34393 
 1.07000   0.36448 
 1.08000   0.38354 
 1.09000   0.40125 
 1.10000   0.41771 
 1.11000   0.43300 
 1.12000   0.44719 
 1.13000   0.46033 
 1.14000   0.47245 
 1.15000   0.48358 
 1.16000   0.49375 
 1.17000   0.50297 
 1.18000   0.51127 
 1.19000   0.51865 
 1.20000   0.52513 
 1.21000   0.53071 
 1.22000   0.53540 
 1.23000   0.53920 
 1.24000   0.54212 
 1.25000   0.54417 
 1.26000   0.54533 
 1.27000   0.54562 
 1.28000   0.54503 
 1.29000   0.54357 
 1.30000   0.54123 
 1.31000   0.53801 
 1.32000   0.53390 
 1.33000   0.52891 
 1.34000   0.52302 
 1.35000   0.51624 
 1.36000   0.50855 
 1.37000   0.49993 
 1.38000   0.49039 
 1.39000   0.47989 
 1.40000   0.46843 
 1.41000   0.45597 
 1.42000   0.44248 
 1.43000   0.42792 
 1.44000   0.41224 
 1.45000   0.39536 
 1.46000   0.37721 
 1.47000   0.35766 
 1.48000   0.33655 
 1.49000   0.31366 
 1.50000   0.28868 
 1.51000   0.26110 
 1.52000   0.23014 
 1.53000   0.19433 
 1.54000   0.15045 
 1.55000   0.08738 
 1.55100   0.07839
 1.55200   0.06824
 1.55300   0.05631
 1.55400   0.04106
 1.55514   0.00000
/
\plot
 1.98146   0.00000
 1.98200   0.02839
 1.98300   0.04785
 1.98400   0.06144
 1.98500   0.07254
 1.98600   0.08217
 1.98700   0.09079
 1.98800   0.09867
 1.98900   0.10598
 1.99000   0.11282 
 2.00000   0.16667 
/
\setdashes <2pt>
\plot
 0.00000  -0.50000 
 0.01000  -0.52696 
 0.02000  -0.55339 
 0.03000  -0.57928 
 0.04000  -0.60460 
 0.05000  -0.62932 
 0.06000  -0.65342 
 0.07000  -0.67688 
 0.08000  -0.69966 
 0.09000  -0.72176 
 0.10000  -0.74314 
 0.11000  -0.76380 
 0.12000  -0.78369 
 0.13000  -0.80282 
 0.14000  -0.82115 
 0.15000  -0.83867 
 0.16000  -0.85536 
 0.17000  -0.87121 
 0.18000  -0.88620 
 0.19000  -0.90032 
 0.20000  -0.91355 
 0.21000  -0.92587 
 0.22000  -0.93728 
 0.23000  -0.94777 
 0.24000  -0.95732 
 0.25000  -0.96593 
 0.26000  -0.97358 
 0.27000  -0.98027 
 0.28000  -0.98600 
 0.29000  -0.99075 
 0.30000  -0.99452 
 0.31000  -0.99731 
 0.32000  -0.99912 
 0.33000  -0.99995 
 0.34000  -0.99978 
 0.35000  -0.99863 
 0.36000  -0.99649 
 0.37000  -0.99337 
 0.38000  -0.98927 
 0.39000  -0.98420 
 0.40000  -0.97815 
 0.41000  -0.97113 
 0.42000  -0.96316 
 0.43000  -0.95424 
 0.44000  -0.94438 
 0.45000  -0.93358 
 0.46000  -0.92186 
 0.47000  -0.90924 
 0.48000  -0.89571 
 0.49000  -0.88130 
 0.50000  -0.86603 
 0.51000  -0.84989 
 0.52000  -0.83292 
 0.53000  -0.81513 
 0.54000  -0.79653 
 0.55000  -0.77715 
 0.55000  -0.77715 
/
\plot
 0.99000   0.47255 
 1.00000   0.50000 
 1.01000   0.52696 
 1.02000   0.55339 
 1.03000   0.57928 
 1.04000   0.60460 
 1.05000   0.62932 
 1.06000   0.65342 
 1.07000   0.67688 
 1.08000   0.69966 
 1.09000   0.72176 
 1.10000   0.74314 
 1.11000   0.76380 
 1.12000   0.78369 
 1.13000   0.80282 
 1.14000   0.82115 
 1.15000   0.83867 
 1.16000   0.85536 
 1.17000   0.87121 
 1.18000   0.88620 
 1.19000   0.90032 
 1.20000   0.91355 
 1.21000   0.92587 
 1.22000   0.93728 
 1.23000   0.94777 
 1.24000   0.95732 
 1.25000   0.96593 
 1.26000   0.97358 
 1.27000   0.98027 
 1.28000   0.98600 
 1.29000   0.99075 
 1.30000   0.99452 
 1.31000   0.99731 
 1.32000   0.99912 
 1.33000   0.99995 
 1.34000   0.99978 
 1.35000   0.99863 
 1.36000   0.99649 
 1.37000   0.99337 
 1.38000   0.98927 
 1.39000   0.98420 
 1.40000   0.97815 
 1.41000   0.97113 
 1.42000   0.96316 
 1.43000   0.95424 
 1.44000   0.94438 
 1.45000   0.93358 
 1.46000   0.92186 
 1.47000   0.90924 
 1.48000   0.89571 
 1.49000   0.88130 
 1.50000   0.86603 
 1.51000   0.84989 
 1.52000   0.83292 
 1.53000   0.81513 
 1.54000   0.79653 
 1.55000   0.77715 
/
\plot
 1.99000  -0.47255 
 2.00000  -0.50000 
 2.00000  -0.50000 
/
}}

\def\plotc{\softplot{%
\put {(c)} <18pt,-15pt> at 0.0 2.0
\put {$\alpha(T+\bar T)=1.0$} <0pt,-15pt> at 1.0 2.0
\put {$M$} at 0.87  0.60
\put {$m$} at 0.87  0.20
\put {$A$} at 0.87 -0.50
\put {$M$} at 1.87 -0.60
\put {$m$} at 1.87  0.20
\put {$A$} at 1.87  0.50
}{
\setdots <2pt>
\plot
 0.00000   0.25000 
 0.01000   0.29068 
 0.02000   0.33107 
 0.03000   0.37114 
 0.04000   0.41084 
 0.05000   0.45014 
 0.06000   0.48899 
 0.07000   0.52736 
 0.08000   0.56520 
 0.09000   0.60249 
 0.10000   0.63919 
 0.11000   0.67525 
 0.12000   0.71065 
 0.13000   0.74535 
 0.14000   0.77931 
 0.15000   0.81250 
 0.16000   0.84489 
 0.17000   0.87645 
 0.18000   0.90714 
 0.19000   0.93694 
 0.20000   0.96581 
 0.21000   0.99373 
 0.22000   1.02067 
 0.23000   1.04660 
 0.24000   1.07149 
 0.25000   1.09534 
 0.26000   1.11809 
 0.27000   1.13975 
 0.28000   1.16028 
 0.29000   1.17967 
 0.30000   1.19789 
 0.31000   1.21493 
 0.32000   1.23077 
 0.33000   1.24540 
 0.34000   1.25879 
 0.35000   1.27095 
 0.36000   1.28185 
 0.37000   1.29149 
 0.38000   1.29985 
 0.39000   1.30692 
 0.40000   1.31271 
 0.41000   1.31721 
 0.42000   1.32040 
 0.43000   1.32229 
 0.44000   1.32287 
 0.45000   1.32215 
 0.46000   1.32013 
 0.47000   1.31680 
 0.48000   1.31217 
 0.49000   1.30625 
 0.50000   1.29904 
 0.51000   1.29054 
 0.52000   1.28078 
 0.53000   1.26975 
 0.54000   1.25746 
 0.55000   1.24394 
 0.56000   1.22918 
 0.57000   1.21322 
 0.58000   1.19605 
 0.59000   1.17771 
 0.60000   1.15820 
 0.61000   1.13756 
 0.62000   1.11578 
 0.63000   1.09291 
 0.64000   1.06896 
 0.65000   1.04395 
 0.66000   1.01792 
 0.67000   0.99088 
 0.68000   0.96286 
 0.69000   0.93389 
 0.70000   0.90400 
 0.71000   0.87321 
 0.72000   0.84157 
 0.73000   0.80909 
 0.74000   0.77582 
 0.75000   0.74178 
 0.76000   0.70701 
 0.77000   0.67154 
 0.78000   0.63541 
 0.79000   0.59865 
 0.80000   0.56130 
/
\plot
 0.99000  -0.20907 
 1.00000  -0.25000 
 1.01000  -0.29068 
 1.02000  -0.33107 
 1.03000  -0.37114 
 1.04000  -0.41084 
 1.05000  -0.45014 
 1.06000  -0.48899 
 1.07000  -0.52736 
 1.08000  -0.56520 
 1.09000  -0.60249 
 1.10000  -0.63919 
 1.11000  -0.67525 
 1.12000  -0.71065 
 1.13000  -0.74535 
 1.14000  -0.77931 
 1.15000  -0.81250 
 1.16000  -0.84489 
 1.17000  -0.87645 
 1.18000  -0.90714 
 1.19000  -0.93694 
 1.20000  -0.96581 
 1.21000  -0.99373 
 1.22000  -1.02067 
 1.23000  -1.04660 
 1.24000  -1.07149 
 1.25000  -1.09534 
 1.26000  -1.11809 
 1.27000  -1.13975 
 1.28000  -1.16028 
 1.29000  -1.17967 
 1.30000  -1.19789 
 1.31000  -1.21493 
 1.32000  -1.23077 
 1.33000  -1.24540 
 1.34000  -1.25879 
 1.35000  -1.27095 
 1.36000  -1.28185 
 1.37000  -1.29149 
 1.38000  -1.29985 
 1.39000  -1.30692 
 1.40000  -1.31271 
 1.41000  -1.31721 
 1.42000  -1.32040 
 1.43000  -1.32229 
 1.44000  -1.32287 
 1.45000  -1.32215 
 1.46000  -1.32013 
 1.47000  -1.31680 
 1.48000  -1.31217 
 1.49000  -1.30625 
 1.50000  -1.29904 
 1.51000  -1.29054 
 1.52000  -1.28078 
 1.53000  -1.26975 
 1.54000  -1.25746 
 1.55000  -1.24394 
 1.56000  -1.22918 
 1.57000  -1.21322 
 1.58000  -1.19605 
 1.59000  -1.17771 
 1.60000  -1.15820 
 1.61000  -1.13756 
 1.62000  -1.11578 
 1.63000  -1.09291 
 1.64000  -1.06896 
 1.65000  -1.04395 
 1.66000  -1.01792 
 1.67000  -0.99088 
 1.68000  -0.96286 
 1.69000  -0.93389 
 1.70000  -0.90400 
 1.71000  -0.87321 
 1.72000  -0.84157 
 1.73000  -0.80909 
 1.74000  -0.77582 
 1.75000  -0.74178 
 1.76000  -0.70701 
 1.77000  -0.67154 
 1.78000  -0.63541 
 1.79000  -0.59865 
 1.80000  -0.56130 
 1.80000  -0.56130 
/
\plot
 1.99000   0.20907 
 2.00000   0.25000 
 2.00000   0.25000 
/
\setsolid
\plot
 0.00000   0.10000 
 0.01000   0.14214 
 0.02000   0.17662 
 0.03000   0.20719 
 0.04000   0.23530 
 0.05000   0.26163 
 0.06000   0.28659 
 0.07000   0.31042 
 0.08000   0.33328 
 0.09000   0.35527 
 0.10000   0.37647 
 0.11000   0.39693 
 0.12000   0.41669 
 0.13000   0.43575 
 0.14000   0.45415 
 0.15000   0.47189 
 0.16000   0.48897 
 0.17000   0.50541 
 0.18000   0.52119 
 0.19000   0.53631 
 0.20000   0.55078 
 0.21000   0.56459 
 0.22000   0.57773 
 0.23000   0.59020 
 0.24000   0.60199 
 0.25000   0.61309 
 0.26000   0.62351 
 0.27000   0.63323 
 0.28000   0.64224 
 0.29000   0.65055 
 0.30000   0.65814 
 0.31000   0.66501 
 0.32000   0.67116 
 0.33000   0.67659 
 0.34000   0.68128 
 0.35000   0.68524 
 0.36000   0.68845 
 0.37000   0.69093 
 0.38000   0.69267 
 0.39000   0.69367 
 0.40000   0.69392 
 0.41000   0.69342 
 0.42000   0.69219 
 0.43000   0.69021 
 0.44000   0.68749 
 0.45000   0.68403 
 0.46000   0.67983 
 0.47000   0.67490 
 0.48000   0.66924 
 0.49000   0.66285 
 0.50000   0.65574 
 0.51000   0.64792 
 0.52000   0.63938 
 0.53000   0.63014 
 0.54000   0.62019 
 0.55000   0.60955 
 0.56000   0.59822 
 0.57000   0.58621 
 0.58000   0.57352 
 0.59000   0.56016 
 0.60000   0.54614 
 0.61000   0.53145 
 0.62000   0.51611 
 0.63000   0.50012 
 0.64000   0.48347 
 0.65000   0.46618 
 0.66000   0.44823 
 0.67000   0.42961 
 0.68000   0.41032 
 0.69000   0.39034 
 0.70000   0.36965 
 0.71000   0.34819 
 0.72000   0.32592 
 0.73000   0.30276 
 0.74000   0.27858 
 0.75000   0.25321 
 0.76000   0.22636 
 0.77000   0.19755 
 0.78000   0.16591 
 0.79000   0.12957 
 0.80000   0.08277 
 0.80100   0.07683
 0.80200   0.07045
 0.80300   0.06350
 0.80400   0.05577
 0.80500   0.04686
 0.80600   0.03593
 0.80700   0.01985
 0.80744   0.00000
/
\plot
 0.98930   0.00000
 0.99000   0.02503 
 1.00000   0.10000 
 1.01000   0.14214 
 1.02000   0.17662 
 1.03000   0.20719 
 1.04000   0.23530 
 1.05000   0.26163 
 1.06000   0.28659 
 1.07000   0.31042 
 1.08000   0.33328 
 1.09000   0.35527 
 1.10000   0.37647 
 1.11000   0.39693 
 1.12000   0.41669 
 1.13000   0.43575 
 1.14000   0.45415 
 1.15000   0.47189 
 1.16000   0.48897 
 1.17000   0.50541 
 1.18000   0.52119 
 1.19000   0.53631 
 1.20000   0.55078 
 1.21000   0.56459 
 1.22000   0.57773 
 1.23000   0.59020 
 1.24000   0.60199 
 1.25000   0.61309 
 1.26000   0.62351 
 1.27000   0.63323 
 1.28000   0.64224 
 1.29000   0.65055 
 1.30000   0.65814 
 1.31000   0.66501 
 1.32000   0.67116 
 1.33000   0.67659 
 1.34000   0.68128 
 1.35000   0.68524 
 1.36000   0.68845 
 1.37000   0.69093 
 1.38000   0.69267 
 1.39000   0.69367 
 1.40000   0.69392 
 1.41000   0.69342 
 1.42000   0.69219 
 1.43000   0.69021 
 1.44000   0.68749 
 1.45000   0.68403 
 1.46000   0.67983 
 1.47000   0.67490 
 1.48000   0.66924 
 1.49000   0.66285 
 1.50000   0.65574 
 1.51000   0.64792 
 1.52000   0.63938 
 1.53000   0.63014 
 1.54000   0.62019 
 1.55000   0.60955 
 1.56000   0.59822 
 1.57000   0.58621 
 1.58000   0.57352 
 1.59000   0.56016 
 1.60000   0.54614 
 1.61000   0.53145 
 1.62000   0.51611 
 1.63000   0.50012 
 1.64000   0.48347 
 1.65000   0.46618 
 1.66000   0.44823 
 1.67000   0.42961 
 1.68000   0.41032 
 1.69000   0.39034 
 1.70000   0.36965 
 1.71000   0.34819 
 1.72000   0.32592 
 1.73000   0.30276 
 1.74000   0.27858 
 1.75000   0.25321 
 1.76000   0.22636 
 1.77000   0.19755 
 1.78000   0.16591 
 1.79000   0.12957 
 1.80000   0.08277 
 1.80000   0.08277 
 1.80100   0.07683
 1.80200   0.07045
 1.80300   0.06350
 1.80400   0.05577
 1.80500   0.04686
 1.80600   0.03593
 1.80700   0.01985
 1.80744   0.00000
 1.80744   0.00000
/
\plot
 1.98930   0.00000
 1.99000   0.02503 
 2.00000   0.10000 
/
\setdashes <2pt>
\plot
 0.00000  -0.30000 
 0.01000  -0.33794 
 0.02000  -0.37554 
 0.03000  -0.41277 
 0.04000  -0.44959 
 0.05000  -0.48597 
 0.06000  -0.52187 
 0.07000  -0.55726 
 0.08000  -0.59210 
 0.09000  -0.62635 
 0.10000  -0.65998 
 0.11000  -0.69296 
 0.12000  -0.72526 
 0.13000  -0.75684 
 0.14000  -0.78768 
 0.15000  -0.81774 
 0.16000  -0.84699 
 0.17000  -0.87540 
 0.18000  -0.90295 
 0.19000  -0.92961 
 0.20000  -0.95536 
 0.21000  -0.98016 
 0.22000  -1.00399 
 0.23000  -1.02683 
 0.24000  -1.04866 
 0.25000  -1.06945 
 0.26000  -1.08919 
 0.27000  -1.10785 
 0.28000  -1.12542 
 0.29000  -1.14188 
 0.30000  -1.15722 
 0.31000  -1.17141 
 0.32000  -1.18444 
 0.33000  -1.19631 
 0.34000  -1.20699 
 0.35000  -1.21649 
 0.36000  -1.22478 
 0.37000  -1.23186 
 0.38000  -1.23773 
 0.39000  -1.24238 
 0.40000  -1.24580 
 0.41000  -1.24799 
 0.42000  -1.24895 
 0.43000  -1.24868 
 0.44000  -1.24717 
 0.45000  -1.24444 
 0.46000  -1.24048 
 0.47000  -1.23529 
 0.48000  -1.22888 
 0.49000  -1.22126 
 0.50000  -1.21244 
 0.51000  -1.20241 
 0.52000  -1.19121 
 0.53000  -1.17882 
 0.54000  -1.16528 
 0.55000  -1.15058 
 0.56000  -1.13475 
 0.57000  -1.11779 
 0.58000  -1.09974 
 0.59000  -1.08060 
 0.60000  -1.06039 
 0.61000  -1.03914 
 0.62000  -1.01686 
 0.63000  -0.99357 
 0.64000  -0.96931 
 0.65000  -0.94409 
 0.66000  -0.91794 
 0.67000  -0.89088 
 0.68000  -0.86295 
 0.69000  -0.83416 
 0.70000  -0.80455 
 0.71000  -0.77414 
 0.72000  -0.74297 
 0.73000  -0.71107 
 0.74000  -0.67846 
 0.75000  -0.64519 
 0.76000  -0.61128 
 0.77000  -0.57676 
 0.78000  -0.54168 
 0.79000  -0.50606 
 0.80000  -0.46995 
/
\plot
 0.99000   0.26177 
 1.00000   0.30000 
 1.01000   0.33794 
 1.02000   0.37554 
 1.03000   0.41277 
 1.04000   0.44959 
 1.05000   0.48597 
 1.06000   0.52187 
 1.07000   0.55726 
 1.08000   0.59210 
 1.09000   0.62635 
 1.10000   0.65998 
 1.11000   0.69296 
 1.12000   0.72526 
 1.13000   0.75684 
 1.14000   0.78768 
 1.15000   0.81774 
 1.16000   0.84699 
 1.17000   0.87540 
 1.18000   0.90295 
 1.19000   0.92961 
 1.20000   0.95536 
 1.21000   0.98016 
 1.22000   1.00399 
 1.23000   1.02683 
 1.24000   1.04866 
 1.25000   1.06945 
 1.26000   1.08919 
 1.27000   1.10785 
 1.28000   1.12542 
 1.29000   1.14188 
 1.30000   1.15722 
 1.31000   1.17141 
 1.32000   1.18444 
 1.33000   1.19631 
 1.34000   1.20699 
 1.35000   1.21649 
 1.36000   1.22478 
 1.37000   1.23186 
 1.38000   1.23773 
 1.39000   1.24238 
 1.40000   1.24580 
 1.41000   1.24799 
 1.42000   1.24895 
 1.43000   1.24868 
 1.44000   1.24717 
 1.45000   1.24444 
 1.46000   1.24048 
 1.47000   1.23529 
 1.48000   1.22888 
 1.49000   1.22126 
 1.50000   1.21244 
 1.51000   1.20241 
 1.52000   1.19121 
 1.53000   1.17882 
 1.54000   1.16528 
 1.55000   1.15058 
 1.56000   1.13475 
 1.57000   1.11779 
 1.58000   1.09974 
 1.59000   1.08060 
 1.60000   1.06039 
 1.61000   1.03914 
 1.62000   1.01686 
 1.63000   0.99357 
 1.64000   0.96931 
 1.65000   0.94409 
 1.66000   0.91794 
 1.67000   0.89088 
 1.68000   0.86295 
 1.69000   0.83416 
 1.70000   0.80455 
 1.71000   0.77414 
 1.72000   0.74297 
 1.73000   0.71107 
 1.74000   0.67846 
 1.75000   0.64519 
 1.76000   0.61128 
 1.77000   0.57676 
 1.78000   0.54168 
 1.79000   0.50606 
 1.80000   0.46995 
 1.80000   0.46995 
/
\plot
 1.99000  -0.26177 
 2.00000  -0.30000 
 2.00000  -0.30000 
/
}}

\def\plotd{\softplot{%
\put {(d)} <18pt,-15pt> at 0.0 2.0
\put {$\alpha(T+\bar T)=0.5$} <0pt,-15pt> at 1.0 2.0
\put {$M$} at 0.50  1.30
\put {$m$} at 0.50  0.70
\put {$A$} at 0.50 -1.30
\put {$M$} at 1.50 -1.30
\put {$m$} at 1.50  0.70
\put {$A$} at 1.50  1.30
}{
\setdots <2pt>
\plot
 0.00000   0.12500 
 0.01000   0.17254 
 0.02000   0.21992 
 0.03000   0.26707 
 0.04000   0.31396 
 0.05000   0.36054 
 0.06000   0.40677 
 0.07000   0.45260 
 0.08000   0.49797 
 0.09000   0.54286 
 0.10000   0.58721 
 0.11000   0.63098 
 0.12000   0.67413 
 0.13000   0.71661 
 0.14000   0.75839 
 0.15000   0.79942 
 0.16000   0.83966 
 0.17000   0.87907 
 0.18000   0.91761 
 0.19000   0.95525 
 0.20000   0.99194 
 0.21000   1.02766 
 0.22000   1.06236 
 0.23000   1.09601 
 0.24000   1.12858 
 0.25000   1.16004 
 0.26000   1.19035 
 0.27000   1.21949 
 0.28000   1.24743 
 0.29000   1.27413 
 0.30000   1.29957 
 0.31000   1.32374 
 0.32000   1.34659 
 0.33000   1.36812 
 0.34000   1.38830 
 0.35000   1.40711 
 0.36000   1.42453 
 0.37000   1.44054 
 0.38000   1.45513 
 0.39000   1.46829 
 0.40000   1.48000 
 0.41000   1.49024 
 0.42000   1.49902 
 0.43000   1.50631 
 0.44000   1.51212 
 0.45000   1.51644 
 0.46000   1.51926 
 0.47000   1.52058 
 0.48000   1.52040 
 0.49000   1.51872 
 0.50000   1.51554 
 0.51000   1.51087 
 0.52000   1.50471 
 0.53000   1.49705 
 0.54000   1.48793 
 0.55000   1.47733 
 0.56000   1.46528 
 0.57000   1.45178 
 0.58000   1.43684 
 0.59000   1.42049 
 0.60000   1.40274 
 0.61000   1.38360 
 0.62000   1.36310 
 0.63000   1.34125 
 0.64000   1.31808 
 0.65000   1.29361 
 0.66000   1.26786 
 0.67000   1.24086 
 0.68000   1.21264 
 0.69000   1.18322 
 0.70000   1.15263 
 0.71000   1.12090 
 0.72000   1.08807 
 0.73000   1.05416 
 0.74000   1.01922 
 0.75000   0.98326 
 0.76000   0.94634 
 0.77000   0.90848 
 0.78000   0.86973 
 0.79000   0.83012 
 0.80000   0.78969 
 0.81000   0.74848 
 0.82000   0.70653 
 0.83000   0.66388 
 0.84000   0.62058 
 0.85000   0.57667 
 0.86000   0.53218 
 0.87000   0.48718 
 0.88000   0.44169 
 0.89000   0.39576 
 0.90000   0.34945 
 0.91000   0.30279 
 0.92000   0.25583 
 0.93000   0.20862 
 0.94000   0.16120 
/
\plot
 1.00000  -0.12500 
 1.01000  -0.17254 
 1.02000  -0.21992 
 1.03000  -0.26707 
 1.04000  -0.31396 
 1.05000  -0.36054 
 1.06000  -0.40677 
 1.07000  -0.45260 
 1.08000  -0.49797 
 1.09000  -0.54286 
 1.10000  -0.58721 
 1.11000  -0.63098 
 1.12000  -0.67413 
 1.13000  -0.71661 
 1.14000  -0.75839 
 1.15000  -0.79942 
 1.16000  -0.83966 
 1.17000  -0.87907 
 1.18000  -0.91761 
 1.19000  -0.95525 
 1.20000  -0.99194 
 1.21000  -1.02766 
 1.22000  -1.06236 
 1.23000  -1.09601 
 1.24000  -1.12858 
 1.25000  -1.16004 
 1.26000  -1.19035 
 1.27000  -1.21949 
 1.28000  -1.24743 
 1.29000  -1.27413 
 1.30000  -1.29957 
 1.31000  -1.32374 
 1.32000  -1.34659 
 1.33000  -1.36812 
 1.34000  -1.38830 
 1.35000  -1.40711 
 1.36000  -1.42453 
 1.37000  -1.44054 
 1.38000  -1.45513 
 1.39000  -1.46829 
 1.40000  -1.48000 
 1.41000  -1.49024 
 1.42000  -1.49902 
 1.43000  -1.50631 
 1.44000  -1.51212 
 1.45000  -1.51644 
 1.46000  -1.51926 
 1.47000  -1.52058 
 1.48000  -1.52040 
 1.49000  -1.51872 
 1.50000  -1.51554 
 1.51000  -1.51087 
 1.52000  -1.50471 
 1.53000  -1.49705 
 1.54000  -1.48793 
 1.55000  -1.47733 
 1.56000  -1.46528 
 1.57000  -1.45178 
 1.58000  -1.43684 
 1.59000  -1.42049 
 1.60000  -1.40274 
 1.61000  -1.38360 
 1.62000  -1.36310 
 1.63000  -1.34125 
 1.64000  -1.31808 
 1.65000  -1.29361 
 1.66000  -1.26786 
 1.67000  -1.24086 
 1.68000  -1.21264 
 1.69000  -1.18322 
 1.70000  -1.15263 
 1.71000  -1.12090 
 1.72000  -1.08807 
 1.73000  -1.05416 
 1.74000  -1.01922 
 1.75000  -0.98326 
 1.76000  -0.94634 
 1.77000  -0.90848 
 1.78000  -0.86973 
 1.79000  -0.83012 
 1.80000  -0.78969 
 1.81000  -0.74848 
 1.82000  -0.70653 
 1.83000  -0.66388 
 1.84000  -0.62058 
 1.85000  -0.57667 
 1.86000  -0.53218 
 1.87000  -0.48718 
 1.88000  -0.44169 
 1.89000  -0.39576 
 1.90000  -0.34945 
 1.91000  -0.30279 
 1.92000  -0.25583 
 1.93000  -0.20862 
 1.94000  -0.16120 
/
\setsolid
\plot
 0.00000   0.04545 
 0.01000   0.08664 
 0.02000   0.11987 
 0.03000   0.15042 
 0.04000   0.17957 
 0.05000   0.20782 
 0.06000   0.23540 
 0.07000   0.26243 
 0.08000   0.28897 
 0.09000   0.31506 
 0.10000   0.34071 
 0.11000   0.36591 
 0.12000   0.39068 
 0.13000   0.41498 
 0.14000   0.43882 
 0.15000   0.46218 
 0.16000   0.48504 
 0.17000   0.50739 
 0.18000   0.52921 
 0.19000   0.55047 
 0.20000   0.57117 
 0.21000   0.59128 
 0.22000   0.61079 
 0.23000   0.62968 
 0.24000   0.64793 
 0.25000   0.66552 
 0.26000   0.68245 
 0.27000   0.69869 
 0.28000   0.71422 
 0.29000   0.72904 
 0.30000   0.74314 
 0.31000   0.75648 
 0.32000   0.76907 
 0.33000   0.78090 
 0.34000   0.79194 
 0.35000   0.80220 
 0.36000   0.81165 
 0.37000   0.82030 
 0.38000   0.82813 
 0.39000   0.83513 
 0.40000   0.84131 
 0.41000   0.84665 
 0.42000   0.85114 
 0.43000   0.85479 
 0.44000   0.85759 
 0.45000   0.85953 
 0.46000   0.86062 
 0.47000   0.86086 
 0.48000   0.86024 
 0.49000   0.85876 
 0.50000   0.85643 
 0.51000   0.85325 
 0.52000   0.84922 
 0.53000   0.84434 
 0.54000   0.83863 
 0.55000   0.83208 
 0.56000   0.82470 
 0.57000   0.81650 
 0.58000   0.80748 
 0.59000   0.79767 
 0.60000   0.78705 
 0.61000   0.77566 
 0.62000   0.76349 
 0.63000   0.75055 
 0.64000   0.73687 
 0.65000   0.72244 
 0.66000   0.70730 
 0.67000   0.69144 
 0.68000   0.67489 
 0.69000   0.65766 
 0.70000   0.63977 
 0.71000   0.62123 
 0.72000   0.60206 
 0.73000   0.58228 
 0.74000   0.56190 
 0.75000   0.54094 
 0.76000   0.51943 
 0.77000   0.49737 
 0.78000   0.47479 
 0.79000   0.45170 
 0.80000   0.42812 
 0.81000   0.40407 
 0.82000   0.37955 
 0.83000   0.35459 
 0.84000   0.32919 
 0.85000   0.30334 
 0.86000   0.27705 
 0.87000   0.25029 
 0.88000   0.22302 
 0.89000   0.19516 
 0.90000   0.16654 
 0.91000   0.13685 
 0.92000   0.10534 
 0.93000   0.06981 
 0.93500   0.04793
 0.94000   0.00906 
 0.94020   0.00000
/
\plot
 0.99528   0.00000
 0.99600   0.01712
 0.99700   0.02673
 0.99800   0.03392
 0.99900   0.04001
 1.00000   0.04545 
 1.01000   0.08664 
 1.02000   0.11987 
 1.03000   0.15042 
 1.04000   0.17957 
 1.05000   0.20782 
 1.06000   0.23540 
 1.07000   0.26243 
 1.08000   0.28897 
 1.09000   0.31506 
 1.10000   0.34071 
 1.11000   0.36591 
 1.12000   0.39068 
 1.13000   0.41498 
 1.14000   0.43882 
 1.15000   0.46218 
 1.16000   0.48504 
 1.17000   0.50739 
 1.18000   0.52921 
 1.19000   0.55047 
 1.20000   0.57117 
 1.21000   0.59128 
 1.22000   0.61079 
 1.23000   0.62968 
 1.24000   0.64793 
 1.25000   0.66552 
 1.26000   0.68245 
 1.27000   0.69869 
 1.28000   0.71422 
 1.29000   0.72904 
 1.30000   0.74314 
 1.31000   0.75648 
 1.32000   0.76907 
 1.33000   0.78090 
 1.34000   0.79194 
 1.35000   0.80220 
 1.36000   0.81165 
 1.37000   0.82030 
 1.38000   0.82813 
 1.39000   0.83513 
 1.40000   0.84131 
 1.41000   0.84665 
 1.42000   0.85114 
 1.43000   0.85479 
 1.44000   0.85759 
 1.45000   0.85953 
 1.46000   0.86062 
 1.47000   0.86086 
 1.48000   0.86024 
 1.49000   0.85876 
 1.50000   0.85643 
 1.51000   0.85325 
 1.52000   0.84922 
 1.53000   0.84434 
 1.54000   0.83863 
 1.55000   0.83208 
 1.56000   0.82470 
 1.57000   0.81650 
 1.58000   0.80748 
 1.59000   0.79767 
 1.60000   0.78705 
 1.61000   0.77566 
 1.62000   0.76349 
 1.63000   0.75055 
 1.64000   0.73687 
 1.65000   0.72244 
 1.66000   0.70730 
 1.67000   0.69144 
 1.68000   0.67489 
 1.69000   0.65766 
 1.70000   0.63977 
 1.71000   0.62123 
 1.72000   0.60206 
 1.73000   0.58228 
 1.74000   0.56190 
 1.75000   0.54094 
 1.76000   0.51943 
 1.77000   0.49737 
 1.78000   0.47479 
 1.79000   0.45170 
 1.80000   0.42812 
 1.81000   0.40407 
 1.82000   0.37955 
 1.83000   0.35459 
 1.84000   0.32919 
 1.85000   0.30334 
 1.86000   0.27705 
 1.87000   0.25029 
 1.88000   0.22302 
 1.89000   0.19516 
 1.90000   0.16654 
 1.91000   0.13685 
 1.92000   0.10534 
 1.93000   0.06981 
 1.94000   0.00906 
 1.94020   0.00000
/
\plot
 1.99528   0.00000
 1.99600   0.01712
 1.99700   0.02673
 1.99800   0.03392
 1.99900   0.04001
/
\setdashes <2pt>
\plot
 0.00000  -0.13636 
 0.01000  -0.18328 
 0.02000  -0.23002 
 0.03000  -0.27653 
 0.04000  -0.32277 
 0.05000  -0.36869 
 0.06000  -0.41424 
 0.07000  -0.45939 
 0.08000  -0.50409 
 0.09000  -0.54828 
 0.10000  -0.59194 
 0.11000  -0.63501 
 0.12000  -0.67745 
 0.13000  -0.71923 
 0.14000  -0.76029 
 0.15000  -0.80061 
 0.16000  -0.84013 
 0.17000  -0.87883 
 0.18000  -0.91666 
 0.19000  -0.95358 
 0.20000  -0.98957 
 0.21000  -1.02457 
 0.22000  -1.05857 
 0.23000  -1.09152 
 0.24000  -1.12339 
 0.25000  -1.15416 
 0.26000  -1.18378 
 0.27000  -1.21224 
 0.28000  -1.23950 
 0.29000  -1.26554 
 0.30000  -1.29033 
 0.31000  -1.31385 
 0.32000  -1.33607 
 0.33000  -1.35697 
 0.34000  -1.37653 
 0.35000  -1.39473 
 0.36000  -1.41156 
 0.37000  -1.42699 
 0.38000  -1.44102 
 0.39000  -1.45362 
 0.40000  -1.46479 
 0.41000  -1.47451 
 0.42000  -1.48278 
 0.43000  -1.48958 
 0.44000  -1.49492 
 0.45000  -1.49878 
 0.46000  -1.50116 
 0.47000  -1.50206 
 0.48000  -1.50147 
 0.49000  -1.49941 
 0.50000  -1.49586 
 0.51000  -1.49084 
 0.52000  -1.48435 
 0.53000  -1.47639 
 0.54000  -1.46698 
 0.55000  -1.45611 
 0.56000  -1.44381 
 0.57000  -1.43009 
 0.58000  -1.41495 
 0.59000  -1.39842 
 0.60000  -1.38051 
 0.61000  -1.36124 
 0.62000  -1.34062 
 0.63000  -1.31868 
 0.64000  -1.29544 
 0.65000  -1.27092 
 0.66000  -1.24514 
 0.67000  -1.21814 
 0.68000  -1.18993 
 0.69000  -1.16055 
 0.70000  -1.13003 
 0.71000  -1.09838 
 0.72000  -1.06566 
 0.73000  -1.03188 
 0.74000  -0.99709 
 0.75000  -0.96131 
 0.76000  -0.92458 
 0.77000  -0.88694 
 0.78000  -0.84843 
 0.79000  -0.80908 
 0.80000  -0.76893 
 0.81000  -0.72802 
 0.82000  -0.68639 
 0.83000  -0.64408 
 0.84000  -0.60114 
 0.85000  -0.55761 
 0.86000  -0.51352 
 0.87000  -0.46893 
 0.88000  -0.42388 
 0.89000  -0.37840 
 0.90000  -0.33256 
 0.91000  -0.28638 
 0.92000  -0.23993 
 0.93000  -0.19323 
 0.94000  -0.14635 
/
\plot
 1.00000   0.13636 
 1.01000   0.18328 
 1.02000   0.23002 
 1.03000   0.27653 
 1.04000   0.32277 
 1.05000   0.36869 
 1.06000   0.41424 
 1.07000   0.45939 
 1.08000   0.50409 
 1.09000   0.54828 
 1.10000   0.59194 
 1.11000   0.63501 
 1.12000   0.67745 
 1.13000   0.71923 
 1.14000   0.76029 
 1.15000   0.80061 
 1.16000   0.84013 
 1.17000   0.87883 
 1.18000   0.91666 
 1.19000   0.95358 
 1.20000   0.98957 
 1.21000   1.02457 
 1.22000   1.05857 
 1.23000   1.09152 
 1.24000   1.12339 
 1.25000   1.15416 
 1.26000   1.18378 
 1.27000   1.21224 
 1.28000   1.23950 
 1.29000   1.26554 
 1.30000   1.29033 
 1.31000   1.31385 
 1.32000   1.33607 
 1.33000   1.35697 
 1.34000   1.37653 
 1.35000   1.39473 
 1.36000   1.41156 
 1.37000   1.42699 
 1.38000   1.44102 
 1.39000   1.45362 
 1.40000   1.46479 
 1.41000   1.47451 
 1.42000   1.48278 
 1.43000   1.48958 
 1.44000   1.49492 
 1.45000   1.49878 
 1.46000   1.50116 
 1.47000   1.50206 
 1.48000   1.50147 
 1.49000   1.49941 
 1.50000   1.49586 
 1.51000   1.49084 
 1.52000   1.48435 
 1.53000   1.47639 
 1.54000   1.46698 
 1.55000   1.45611 
 1.56000   1.44381 
 1.57000   1.43009 
 1.58000   1.41495 
 1.59000   1.39842 
 1.60000   1.38051 
 1.61000   1.36124 
 1.62000   1.34062 
 1.63000   1.31868 
 1.64000   1.29544 
 1.65000   1.27092 
 1.66000   1.24514 
 1.67000   1.21814 
 1.68000   1.18993 
 1.69000   1.16055 
 1.70000   1.13003 
 1.71000   1.09838 
 1.72000   1.06566 
 1.73000   1.03188 
 1.74000   0.99709 
 1.75000   0.96131 
 1.76000   0.92458 
 1.77000   0.88694 
 1.78000   0.84843 
 1.79000   0.80908 
 1.80000   0.76893 
 1.81000   0.72802 
 1.82000   0.68639 
 1.83000   0.64408 
 1.84000   0.60114 
 1.85000   0.55761 
 1.86000   0.51352 
 1.87000   0.46893 
 1.88000   0.42388 
 1.89000   0.37840 
 1.90000   0.33256 
 1.91000   0.28638 
 1.92000   0.23993 
 1.93000   0.19323 
 1.94000   0.14635 
/
}}

\def\plotw{\softplot{%
\put {$M$} at 0.50  1.50
\put {$m$} at 0.50  0.80
\put {$A$} at 0.50 -1.50
\put {$M$} at 1.50 -1.50
\put {$m$} at 1.50  0.80
\put {$A$} at 1.50  1.50
}{
\setdots <2pt>
\plot
 0.00000   0.00003 
 0.01000   0.05443 
 0.02000   0.10878 
 0.03000   0.16302 
 0.04000   0.21710 
 0.05000   0.27097 
 0.06000   0.32457 
 0.07000   0.37785 
 0.08000   0.43076 
 0.09000   0.48324 
 0.10000   0.53524 
 0.11000   0.58672 
 0.12000   0.63762 
 0.13000   0.68789 
 0.14000   0.73748 
 0.15000   0.78634 
 0.16000   0.83442 
 0.17000   0.88169 
 0.18000   0.92808 
 0.19000   0.97355 
 0.20000   1.01807 
 0.21000   1.06158 
 0.22000   1.10404 
 0.23000   1.14542 
 0.24000   1.18566 
 0.25000   1.22473 
 0.26000   1.26260 
 0.27000   1.29921 
 0.28000   1.33455 
 0.29000   1.36857 
 0.30000   1.40124 
 0.31000   1.43252 
 0.32000   1.46240 
 0.33000   1.49082 
 0.34000   1.51778 
 0.35000   1.54324 
 0.36000   1.56718 
 0.37000   1.58957 
 0.38000   1.61039 
 0.39000   1.62962 
 0.40000   1.64724 
 0.41000   1.66324 
 0.42000   1.67760 
 0.43000   1.69030 
 0.44000   1.70133 
 0.45000   1.71069 
 0.46000   1.71835 
 0.47000   1.72432 
 0.48000   1.72859 
 0.49000   1.73115 
 0.50000   1.73201 
 0.51000   1.73115 
 0.52000   1.72859 
 0.53000   1.72432 
 0.54000   1.71835 
 0.55000   1.71068 
 0.56000   1.70132 
 0.57000   1.69029 
 0.58000   1.67759 
 0.59000   1.66323 
 0.60000   1.64723 
 0.61000   1.62960 
 0.62000   1.61037 
 0.63000   1.58955 
 0.64000   1.56716 
 0.65000   1.54322 
 0.66000   1.51776 
 0.67000   1.49080 
 0.68000   1.46237 
 0.69000   1.43250 
 0.70000   1.40121 
 0.71000   1.36854 
 0.72000   1.33452 
 0.73000   1.29918 
 0.74000   1.26256 
 0.75000   1.22470 
 0.76000   1.18562 
 0.77000   1.14538 
 0.78000   1.10400 
 0.79000   1.06154 
 0.80000   1.01803 
 0.81000   0.97351 
 0.82000   0.92803 
 0.83000   0.88164 
 0.84000   0.83438 
 0.85000   0.78629 
 0.86000   0.73743 
 0.87000   0.68784 
 0.88000   0.63757 
 0.89000   0.58667 
 0.90000   0.53520 
 0.91000   0.48319 
 0.92000   0.43071 
 0.93000   0.37780 
 0.94000   0.32452 
 0.95000   0.27092 
 0.96000   0.21705 
 0.97000   0.16297 
 0.98000   0.10873 
 0.99000   0.05438 
 1.00000  -0.00003 
 1.01000  -0.05443 
 1.02000  -0.10878 
 1.03000  -0.16302 
 1.04000  -0.21710 
 1.05000  -0.27097 
 1.06000  -0.32457 
 1.07000  -0.37785 
 1.08000  -0.43076 
 1.09000  -0.48324 
 1.10000  -0.53524 
 1.11000  -0.58672 
 1.12000  -0.63762 
 1.13000  -0.68789 
 1.14000  -0.73748 
 1.15000  -0.78634 
 1.16000  -0.83442 
 1.17000  -0.88169 
 1.18000  -0.92808 
 1.19000  -0.97355 
 1.20000  -1.01807 
 1.21000  -1.06158 
 1.22000  -1.10404 
 1.23000  -1.14542 
 1.24000  -1.18566 
 1.25000  -1.22473 
 1.26000  -1.26260 
 1.27000  -1.29921 
 1.28000  -1.33455 
 1.29000  -1.36857 
 1.30000  -1.40124 
 1.31000  -1.43252 
 1.32000  -1.46240 
 1.33000  -1.49082 
 1.34000  -1.51778 
 1.35000  -1.54324 
 1.36000  -1.56718 
 1.37000  -1.58957 
 1.38000  -1.61039 
 1.39000  -1.62962 
 1.40000  -1.64724 
 1.41000  -1.66324 
 1.42000  -1.67760 
 1.43000  -1.69030 
 1.44000  -1.70133 
 1.45000  -1.71069 
 1.46000  -1.71835 
 1.47000  -1.72432 
 1.48000  -1.72859 
 1.49000  -1.73115 
 1.50000  -1.73201 
 1.51000  -1.73115 
 1.52000  -1.72859 
 1.53000  -1.72432 
 1.54000  -1.71835 
 1.55000  -1.71068 
 1.56000  -1.70132 
 1.57000  -1.69029 
 1.58000  -1.67759 
 1.59000  -1.66323 
 1.60000  -1.64723 
 1.61000  -1.62960 
 1.62000  -1.61037 
 1.63000  -1.58955 
 1.64000  -1.56716 
 1.65000  -1.54322 
 1.66000  -1.51776 
 1.67000  -1.49080 
 1.68000  -1.46237 
 1.69000  -1.43250 
 1.70000  -1.40121 
 1.71000  -1.36854 
 1.72000  -1.33452 
 1.73000  -1.29918 
 1.74000  -1.26256 
 1.75000  -1.22470 
 1.76000  -1.18562 
 1.77000  -1.14538 
 1.78000  -1.10400 
 1.79000  -1.06154 
 1.80000  -1.01803 
 1.81000  -0.97351 
 1.82000  -0.92803 
 1.83000  -0.88164 
 1.84000  -0.83438 
 1.85000  -0.78629 
 1.86000  -0.73743 
 1.87000  -0.68784 
 1.88000  -0.63757 
 1.89000  -0.58667 
 1.90000  -0.53520 
 1.91000  -0.48319 
 1.92000  -0.43071 
 1.93000  -0.37780 
 1.94000  -0.32452 
 1.95000  -0.27092 
 1.96000  -0.21705 
 1.97000  -0.16297 
 1.98000  -0.10873 
 1.99000  -0.05438 
 2.00000   0.00003 
/
\setsolid
\plot
 0.00000   0.00000 
 0.01000   0.03142 
 0.02000   0.06280 
 0.03000   0.09412 
 0.04000   0.12534 
 0.05000   0.15644 
 0.06000   0.18739 
 0.07000   0.21815 
 0.08000   0.24870 
 0.09000   0.27900 
 0.10000   0.30902 
 0.11000   0.33874 
 0.12000   0.36813 
 0.13000   0.39715 
 0.14000   0.42578 
 0.15000   0.45399 
 0.16000   0.48175 
 0.17000   0.50904 
 0.18000   0.53583 
 0.19000   0.56208 
 0.20000   0.58778 
 0.21000   0.61290 
 0.22000   0.63742 
 0.23000   0.66131 
 0.24000   0.68454 
 0.25000   0.70710 
 0.26000   0.72896 
 0.27000   0.75010 
 0.28000   0.77050 
 0.29000   0.79014 
 0.30000   0.80901 
 0.31000   0.82707 
 0.32000   0.84431 
 0.33000   0.86073 
 0.34000   0.87629 
 0.35000   0.89099 
 0.36000   0.90481 
 0.37000   0.91774 
 0.38000   0.92976 
 0.39000   0.94086 
 0.40000   0.95104 
 0.41000   0.96027 
 0.42000   0.96856 
 0.43000   0.97590 
 0.44000   0.98227 
 0.45000   0.98767 
 0.46000   0.99209 
 0.47000   0.99554 
 0.48000   0.99800 
 0.49000   0.99948 
 0.50000   0.99997 
 0.51000   0.99948 
 0.52000   0.99800 
 0.53000   0.99554 
 0.54000   0.99209 
 0.55000   0.98766 
 0.56000   0.98226 
 0.57000   0.97589 
 0.58000   0.96856 
 0.59000   0.96027 
 0.60000   0.95103 
 0.61000   0.94085 
 0.62000   0.92975 
 0.63000   0.91773 
 0.64000   0.90480 
 0.65000   0.89098 
 0.66000   0.87628 
 0.67000   0.86071 
 0.68000   0.84430 
 0.69000   0.82705 
 0.70000   0.80899 
 0.71000   0.79013 
 0.72000   0.77048 
 0.73000   0.75008 
 0.74000   0.72894 
 0.75000   0.70708 
 0.76000   0.68452 
 0.77000   0.66128 
 0.78000   0.63740 
 0.79000   0.61288 
 0.80000   0.58776 
 0.81000   0.56206 
 0.82000   0.53580 
 0.83000   0.50902 
 0.84000   0.48173 
 0.85000   0.45397 
 0.86000   0.42576 
 0.87000   0.39712 
 0.88000   0.36810 
 0.89000   0.33872 
 0.90000   0.30900 
 0.91000   0.27897 
 0.92000   0.24867 
 0.93000   0.21812 
 0.94000   0.18736 
 0.95000   0.15642 
 0.96000   0.12532 
 0.97000   0.09409 
 0.98000   0.06277 
 0.99000   0.03140 
 1.00000   0.00000 
 1.01000   0.03142 
 1.02000   0.06280 
 1.03000   0.09412 
 1.04000   0.12534 
 1.05000   0.15644 
 1.06000   0.18739 
 1.07000   0.21815 
 1.08000   0.24870 
 1.09000   0.27900 
 1.10000   0.30902 
 1.11000   0.33874 
 1.12000   0.36813 
 1.13000   0.39715 
 1.14000   0.42578 
 1.15000   0.45399 
 1.16000   0.48175 
 1.17000   0.50904 
 1.18000   0.53583 
 1.19000   0.56208 
 1.20000   0.58778 
 1.21000   0.61290 
 1.22000   0.63742 
 1.23000   0.66131 
 1.24000   0.68454 
 1.25000   0.70710 
 1.26000   0.72896 
 1.27000   0.75010 
 1.28000   0.77050 
 1.29000   0.79014 
 1.30000   0.80901 
 1.31000   0.82707 
 1.32000   0.84431 
 1.33000   0.86073 
 1.34000   0.87629 
 1.35000   0.89099 
 1.36000   0.90481 
 1.37000   0.91774 
 1.38000   0.92976 
 1.39000   0.94086 
 1.40000   0.95104 
 1.41000   0.96027 
 1.42000   0.96856 
 1.43000   0.97590 
 1.44000   0.98227 
 1.45000   0.98767 
 1.46000   0.99209 
 1.47000   0.99554 
 1.48000   0.99800 
 1.49000   0.99948 
 1.50000   0.99997 
 1.51000   0.99948 
 1.52000   0.99800 
 1.53000   0.99554 
 1.54000   0.99209 
 1.55000   0.98766 
 1.56000   0.98226 
 1.57000   0.97589 
 1.58000   0.96856 
 1.59000   0.96027 
 1.60000   0.95103 
 1.61000   0.94085 
 1.62000   0.92975 
 1.63000   0.91773 
 1.64000   0.90480 
 1.65000   0.89098 
 1.66000   0.87628 
 1.67000   0.86071 
 1.68000   0.84430 
 1.69000   0.82705 
 1.70000   0.80899 
 1.71000   0.79013 
 1.72000   0.77048 
 1.73000   0.75008 
 1.74000   0.72894 
 1.75000   0.70708 
 1.76000   0.68452 
 1.77000   0.66128 
 1.78000   0.63740 
 1.79000   0.61288 
 1.80000   0.58776 
 1.81000   0.56206 
 1.82000   0.53580 
 1.83000   0.50902 
 1.84000   0.48173 
 1.85000   0.45397 
 1.86000   0.42576 
 1.87000   0.39712 
 1.88000   0.36810 
 1.89000   0.33872 
 1.90000   0.30900 
 1.91000   0.27897 
 1.92000   0.24867 
 1.93000   0.21812 
 1.94000   0.18736 
 1.95000   0.15642 
 1.96000   0.12532 
 1.97000   0.09409 
 1.98000   0.06277 
 1.99000   0.03140 
 2.00000   0.00000 
/
\setdashes <2pt>
\plot
 0.00000  -0.00003 
 0.01000  -0.05443 
 0.02000  -0.10878 
 0.03000  -0.16302 
 0.04000  -0.21710 
 0.05000  -0.27097 
 0.06000  -0.32457 
 0.07000  -0.37785 
 0.08000  -0.43076 
 0.09000  -0.48324 
 0.10000  -0.53524 
 0.11000  -0.58672 
 0.12000  -0.63762 
 0.13000  -0.68789 
 0.14000  -0.73748 
 0.15000  -0.78634 
 0.16000  -0.83442 
 0.17000  -0.88169 
 0.18000  -0.92808 
 0.19000  -0.97355 
 0.20000  -1.01807 
 0.21000  -1.06158 
 0.22000  -1.10404 
 0.23000  -1.14542 
 0.24000  -1.18566 
 0.25000  -1.22473 
 0.26000  -1.26260 
 0.27000  -1.29921 
 0.28000  -1.33455 
 0.29000  -1.36857 
 0.30000  -1.40124 
 0.31000  -1.43252 
 0.32000  -1.46240 
 0.33000  -1.49082 
 0.34000  -1.51778 
 0.35000  -1.54324 
 0.36000  -1.56718 
 0.37000  -1.58957 
 0.38000  -1.61039 
 0.39000  -1.62962 
 0.40000  -1.64724 
 0.41000  -1.66324 
 0.42000  -1.67760 
 0.43000  -1.69030 
 0.44000  -1.70133 
 0.45000  -1.71069 
 0.46000  -1.71835 
 0.47000  -1.72432 
 0.48000  -1.72859 
 0.49000  -1.73115 
 0.50000  -1.73201 
 0.51000  -1.73115 
 0.52000  -1.72859 
 0.53000  -1.72432 
 0.54000  -1.71835 
 0.55000  -1.71068 
 0.56000  -1.70132 
 0.57000  -1.69029 
 0.58000  -1.67759 
 0.59000  -1.66323 
 0.60000  -1.64723 
 0.61000  -1.62960 
 0.62000  -1.61037 
 0.63000  -1.58955 
 0.64000  -1.56716 
 0.65000  -1.54322 
 0.66000  -1.51776 
 0.67000  -1.49080 
 0.68000  -1.46237 
 0.69000  -1.43250 
 0.70000  -1.40121 
 0.71000  -1.36854 
 0.72000  -1.33452 
 0.73000  -1.29918 
 0.74000  -1.26256 
 0.75000  -1.22470 
 0.76000  -1.18562 
 0.77000  -1.14538 
 0.78000  -1.10400 
 0.79000  -1.06154 
 0.80000  -1.01803 
 0.81000  -0.97351 
 0.82000  -0.92803 
 0.83000  -0.88164 
 0.84000  -0.83438 
 0.85000  -0.78629 
 0.86000  -0.73743 
 0.87000  -0.68784 
 0.88000  -0.63757 
 0.89000  -0.58667 
 0.90000  -0.53520 
 0.91000  -0.48319 
 0.92000  -0.43071 
 0.93000  -0.37780 
 0.94000  -0.32452 
 0.95000  -0.27092 
 0.96000  -0.21705 
 0.97000  -0.16297 
 0.98000  -0.10873 
 0.99000  -0.05438 
 1.00000   0.00003 
 1.01000   0.05443 
 1.02000   0.10878 
 1.03000   0.16302 
 1.04000   0.21710 
 1.05000   0.27097 
 1.06000   0.32457 
 1.07000   0.37785 
 1.08000   0.43076 
 1.09000   0.48324 
 1.10000   0.53524 
 1.11000   0.58672 
 1.12000   0.63762 
 1.13000   0.68789 
 1.14000   0.73748 
 1.15000   0.78634 
 1.16000   0.83442 
 1.17000   0.88169 
 1.18000   0.92808 
 1.19000   0.97355 
 1.20000   1.01807 
 1.21000   1.06158 
 1.22000   1.10404 
 1.23000   1.14542 
 1.24000   1.18566 
 1.25000   1.22473 
 1.26000   1.26260 
 1.27000   1.29921 
 1.28000   1.33455 
 1.29000   1.36857 
 1.30000   1.40124 
 1.31000   1.43252 
 1.32000   1.46240 
 1.33000   1.49082 
 1.34000   1.51778 
 1.35000   1.54324 
 1.36000   1.56718 
 1.37000   1.58957 
 1.38000   1.61039 
 1.39000   1.62962 
 1.40000   1.64724 
 1.41000   1.66324 
 1.42000   1.67760 
 1.43000   1.69030 
 1.44000   1.70133 
 1.45000   1.71069 
 1.46000   1.71835 
 1.47000   1.72432 
 1.48000   1.72859 
 1.49000   1.73115 
 1.50000   1.73201 
 1.51000   1.73115 
 1.52000   1.72859 
 1.53000   1.72432 
 1.54000   1.71835 
 1.55000   1.71068 
 1.56000   1.70132 
 1.57000   1.69029 
 1.58000   1.67759 
 1.59000   1.66323 
 1.60000   1.64723 
 1.61000   1.62960 
 1.62000   1.61037 
 1.63000   1.58955 
 1.64000   1.56716 
 1.65000   1.54322 
 1.66000   1.51776 
 1.67000   1.49080 
 1.68000   1.46237 
 1.69000   1.43250 
 1.70000   1.40121 
 1.71000   1.36854 
 1.72000   1.33452 
 1.73000   1.29918 
 1.74000   1.26256 
 1.75000   1.22470 
 1.76000   1.18562 
 1.77000   1.14538 
 1.78000   1.10400 
 1.79000   1.06154 
 1.80000   1.01803 
 1.81000   0.97351 
 1.82000   0.92803 
 1.83000   0.88164 
 1.84000   0.83438 
 1.85000   0.78629 
 1.86000   0.73743 
 1.87000   0.68784 
 1.88000   0.63757 
 1.89000   0.58667 
 1.90000   0.53520 
 1.91000   0.48319 
 1.92000   0.43071 
 1.93000   0.37780 
 1.94000   0.32452 
 1.95000   0.27092 
 1.96000   0.21705 
 1.97000   0.16297 
 1.98000   0.10873 
 1.99000   0.05438 
 2.00000  -0.00003 
/
}}

%
%

\def\ratioplot{
\beginpicture
\setcoordinatesystem units <150pt,300pt> point at 0 0
\setplotarea x from 0.0 to 2.0, y from 0.0 to 0.7
\inboundscheckon
\linethickness 0.5pt
\axis bottom label {}
      ticks in
      width <0.5pt> length <3.0pt> unlabeled quantity 21
      width <0.5pt> length <6.0pt>
      withvalues {$0$} {$\frac12\pi$} {$\pi$} {$\frac32\pi$} {$2\pi$} /
      at 0 0.5 1.0 1.5 2.0 /
/
\axis left label {\large$\displaystyle\frac{m}{|M|}$}
      ticks in
      width <0.5pt> length <6.0pt>
      withvalues {0} {0.2} {0.4} {0.6} /
      at 0.0 0.2 0.4 0.6 /
/
\axis top label {}
      ticks in
      width <0.5pt> length <3.0pt> unlabeled quantity 21
      width <0.5pt> length <6.0pt>
      at 0 0.5 1.0 1.5 2.0 /
/
\axis right label {}
      ticks in
      width <0.5pt> length <6.0pt>
      at 0.0 0.2 0.4 0.6 /
/
\setplotsymbol ({.})
\setdashpattern <10pt,2pt,2pt,2pt>
\plot
 0.00000   0.57735 
 2.00000   0.57735
/
\setdots <2pt>
   \setquadratic 
\plot
 0.00000   0.36364 
 0.01000   0.50214 
 0.02000   0.54507 
 0.03000   0.56321 
 0.04000   0.57194 
 0.05000   0.57640 
 0.06000   0.57870 
 0.07000   0.57983 
 0.08000   0.58030 
 0.09000   0.58037 
 0.10000   0.58021 
 0.11000   0.57991 
 0.12000   0.57952 
 0.13000   0.57909 
 0.14000   0.57862 
 0.15000   0.57815 
 0.16000   0.57767 
 0.17000   0.57719 
 0.18000   0.57672 
 0.19000   0.57626 
 0.20000   0.57581 
 0.21000   0.57537 
 0.22000   0.57494 
 0.23000   0.57452 
 0.24000   0.57411 
 0.25000   0.57371 
 0.26000   0.57331 
 0.27000   0.57293 
 0.28000   0.57256 
 0.29000   0.57219 
 0.30000   0.57183 
 0.31000   0.57147 
 0.32000   0.57113 
 0.33000   0.57078 
 0.34000   0.57044 
 0.35000   0.57010 
 0.36000   0.56977 
 0.37000   0.56944 
 0.38000   0.56911 
 0.39000   0.56878 
 0.40000   0.56845 
 0.41000   0.56813 
 0.42000   0.56780 
 0.43000   0.56747 
 0.44000   0.56714 
 0.45000   0.56681 
 0.46000   0.56647 
 0.47000   0.56614 
 0.48000   0.56579 
 0.49000   0.56545 
 0.50000   0.56510 
 0.51000   0.56474 
 0.52000   0.56437 
 0.53000   0.56400 
 0.54000   0.56362 
 0.55000   0.56323 
 0.56000   0.56283 
 0.57000   0.56241 
 0.58000   0.56198 
 0.59000   0.56154 
 0.60000   0.56108 
 0.61000   0.56061 
 0.62000   0.56011 
 0.63000   0.55959 
 0.64000   0.55904 
 0.65000   0.55847 
 0.66000   0.55787 
 0.67000   0.55723 
 0.68000   0.55655 
 0.69000   0.55583 
 0.70000   0.55505 
 0.71000   0.55422 
 0.72000   0.55333 
 0.73000   0.55236 
 0.74000   0.55130 
 0.75000   0.55015 
 0.76000   0.54888 
 0.77000   0.54747 
 0.78000   0.54590 
 0.79000   0.54414 
 0.80000   0.54214 
 0.81000   0.53985 
 0.82000   0.53721 
 0.83000   0.53412 
 0.84000   0.53045 
 0.85000   0.52602 
 0.86000   0.52059 
 0.87000   0.51376 
 0.88000   0.50493 
 0.89000   0.49312 
 0.90000   0.47659 
 0.91000   0.45197 
 0.92000   0.41177 
 0.93000   0.33463 
 0.94000   0.05621 
 0.94010   0.02811 
 0.94020   0.00000
/
\plot
 0.99528   0.00000
 0.99600   0.16158
 0.99700   0.24146
 0.99800   0.29375
 0.99900   0.33279
 1.00000   0.36364 
 1.01000   0.50214 
 1.02000   0.54507 
 1.03000   0.56321 
 1.04000   0.57194 
 1.05000   0.57640 
 1.06000   0.57870 
 1.07000   0.57983 
 1.08000   0.58030 
 1.09000   0.58037 
 1.10000   0.58021 
 1.11000   0.57991 
 1.12000   0.57952 
 1.13000   0.57909 
 1.14000   0.57862 
 1.15000   0.57815 
 1.16000   0.57767 
 1.17000   0.57719 
 1.18000   0.57672 
 1.19000   0.57626 
 1.20000   0.57581 
 1.21000   0.57537 
 1.22000   0.57494 
 1.23000   0.57452 
 1.24000   0.57411 
 1.25000   0.57371 
 1.26000   0.57331 
 1.27000   0.57293 
 1.28000   0.57256 
 1.29000   0.57219 
 1.30000   0.57183 
 1.31000   0.57147 
 1.32000   0.57113 
 1.33000   0.57078 
 1.34000   0.57044 
 1.35000   0.57010 
 1.36000   0.56977 
 1.37000   0.56944 
 1.38000   0.56911 
 1.39000   0.56878 
 1.40000   0.56845 
 1.41000   0.56813 
 1.42000   0.56780 
 1.43000   0.56747 
 1.44000   0.56714 
 1.45000   0.56681 
 1.46000   0.56647 
 1.47000   0.56614 
 1.48000   0.56579 
 1.49000   0.56545 
 1.50000   0.56510 
 1.51000   0.56474 
 1.52000   0.56437 
 1.53000   0.56400 
 1.54000   0.56362 
 1.55000   0.56323 
 1.56000   0.56283 
 1.57000   0.56241 
 1.58000   0.56198 
 1.59000   0.56154 
 1.60000   0.56108 
 1.61000   0.56061 
 1.62000   0.56011 
 1.63000   0.55959 
 1.64000   0.55904 
 1.65000   0.55847 
 1.66000   0.55787 
 1.67000   0.55723 
 1.68000   0.55655 
 1.69000   0.55583 
 1.70000   0.55505 
 1.71000   0.55422 
 1.72000   0.55333 
 1.73000   0.55236 
 1.74000   0.55130 
 1.75000   0.55015 
 1.76000   0.54888 
 1.77000   0.54747 
 1.78000   0.54590 
 1.79000   0.54414 
 1.80000   0.54214 
 1.81000   0.53985 
 1.82000   0.53721 
 1.83000   0.53412 
 1.84000   0.53045 
 1.85000   0.52602 
 1.86000   0.52059 
 1.87000   0.51376 
 1.88000   0.50493 
 1.89000   0.49312 
 1.90000   0.47659 
 1.91000   0.45197 
 1.92000   0.41177 
 1.93000   0.33463 
 1.94000   0.05621 
 1.94020   0.00000
/
\setdashes <4pt>
   \setquadratic 
\plot
 0.00000   0.40000 
 0.01000   0.48897 
 0.02000   0.53347 
 0.03000   0.55826 
 0.04000   0.57272 
 0.05000   0.58122 
 0.06000   0.58608 
 0.07000   0.58863 
 0.08000   0.58966 
 0.09000   0.58967 
 0.10000   0.58899 
 0.11000   0.58783 
 0.12000   0.58634 
 0.13000   0.58463 
 0.14000   0.58276 
 0.15000   0.58078 
 0.16000   0.57874 
 0.17000   0.57665 
 0.18000   0.57454 
 0.19000   0.57241 
 0.20000   0.57028 
 0.21000   0.56815 
 0.22000   0.56603 
 0.23000   0.56392 
 0.24000   0.56182 
 0.25000   0.55973 
 0.26000   0.55765 
 0.27000   0.55558 
 0.28000   0.55352 
 0.29000   0.55147 
 0.30000   0.54942 
 0.31000   0.54737 
 0.32000   0.54532 
 0.33000   0.54327 
 0.34000   0.54122 
 0.35000   0.53915 
 0.36000   0.53708 
 0.37000   0.53499 
 0.38000   0.53289 
 0.39000   0.53076 
 0.40000   0.52861 
 0.41000   0.52644 
 0.42000   0.52423 
 0.43000   0.52198 
 0.44000   0.51969 
 0.45000   0.51736 
 0.46000   0.51497 
 0.47000   0.51253 
 0.48000   0.51002 
 0.49000   0.50745 
 0.50000   0.50479 
 0.51000   0.50205 
 0.52000   0.49921 
 0.53000   0.49627 
 0.54000   0.49321 
 0.55000   0.49002 
 0.56000   0.48668 
 0.57000   0.48319 
 0.58000   0.47951 
 0.59000   0.47564 
 0.60000   0.47154 
 0.61000   0.46719 
 0.62000   0.46256 
 0.63000   0.45760 
 0.64000   0.45229 
 0.65000   0.44655 
 0.66000   0.44034 
 0.67000   0.43357 
 0.68000   0.42615 
 0.69000   0.41798 
 0.70000   0.40890 
 0.71000   0.39875 
 0.72000   0.38728 
 0.73000   0.37420 
 0.74000   0.35908 
 0.75000   0.34135 
 0.76000   0.32016 
 0.77000   0.29417 
 0.78000   0.26111 
 0.79000   0.21643 
 0.80000   0.14747 
 0.80100   0.13781
 0.80200   0.12723
 0.80300   0.11546
 0.80400   0.10210
 0.80500   0.08640
 0.80600   0.06670
 0.80700   0.03712
 0.80744   0.00000
/
\plot
 0.98930   0.00000
 0.99000   0.11970 
 1.00000   0.40000 
 1.01000   0.48897 
 1.02000   0.53347 
 1.03000   0.55826 
 1.04000   0.57272 
 1.05000   0.58122 
 1.06000   0.58608 
 1.07000   0.58863 
 1.08000   0.58966 
 1.09000   0.58967 
 1.10000   0.58899 
 1.11000   0.58783 
 1.12000   0.58634 
 1.13000   0.58463 
 1.14000   0.58276 
 1.15000   0.58078 
 1.16000   0.57874 
 1.17000   0.57665 
 1.18000   0.57454 
 1.19000   0.57241 
 1.20000   0.57028 
 1.21000   0.56815 
 1.22000   0.56603 
 1.23000   0.56392 
 1.24000   0.56182 
 1.25000   0.55973 
 1.26000   0.55765 
 1.27000   0.55558 
 1.28000   0.55352 
 1.29000   0.55147 
 1.30000   0.54942 
 1.31000   0.54737 
 1.32000   0.54532 
 1.33000   0.54327 
 1.34000   0.54122 
 1.35000   0.53915 
 1.36000   0.53708 
 1.37000   0.53499 
 1.38000   0.53289 
 1.39000   0.53076 
 1.40000   0.52861 
 1.41000   0.52644 
 1.42000   0.52423 
 1.43000   0.52198 
 1.44000   0.51969 
 1.45000   0.51736 
 1.46000   0.51497 
 1.47000   0.51253 
 1.48000   0.51002 
 1.49000   0.50745 
 1.50000   0.50479 
 1.51000   0.50205 
 1.52000   0.49921 
 1.53000   0.49627 
 1.54000   0.49321 
 1.55000   0.49002 
 1.56000   0.48668 
 1.57000   0.48319 
 1.58000   0.47951 
 1.59000   0.47564 
 1.60000   0.47154 
 1.61000   0.46719 
 1.62000   0.46256 
 1.63000   0.45760 
 1.64000   0.45229 
 1.65000   0.44655 
 1.66000   0.44034 
 1.67000   0.43357 
 1.68000   0.42615 
 1.69000   0.41798 
 1.70000   0.40890 
 1.71000   0.39875 
 1.72000   0.38728 
 1.73000   0.37420 
 1.74000   0.35908 
 1.75000   0.34135 
 1.76000   0.32016 
 1.77000   0.29417 
 1.78000   0.26111 
 1.79000   0.21643 
 1.80000   0.14747 
 1.80100   0.13781
 1.80200   0.12723
 1.80300   0.11546
 1.80400   0.10210
 1.80500   0.08640
 1.80600   0.06670
 1.80700   0.03712
 1.80744   0.00000
/
\plot
 1.98930   0.00000
 1.99000   0.11970 
 2.00000   0.40000 
/
\setdashes <2pt>
   \setquadratic 
\plot
 0.00000   0.44444 
 0.01000   0.50686 
 0.02000   0.54527 
 0.03000   0.56991 
 0.04000   0.58585 
 0.05000   0.59600 
 0.06000   0.60211 
 0.07000   0.60534 
 0.08000   0.60646 
 0.09000   0.60600 
 0.10000   0.60436 
 0.11000   0.60179 
 0.12000   0.59852 
 0.13000   0.59468 
 0.14000   0.59039 
 0.15000   0.58574 
 0.16000   0.58079 
 0.17000   0.57559 
 0.18000   0.57018 
 0.19000   0.56459 
 0.20000   0.55884 
 0.21000   0.55294 
 0.22000   0.54690 
 0.23000   0.54073 
 0.24000   0.53443 
 0.25000   0.52799 
 0.26000   0.52143 
 0.27000   0.51473 
 0.28000   0.50789 
 0.29000   0.50089 
 0.30000   0.49373 
 0.31000   0.48639 
 0.32000   0.47886 
 0.33000   0.47112 
 0.34000   0.46314 
 0.35000   0.45492 
 0.36000   0.44642 
 0.37000   0.43761 
 0.38000   0.42845 
 0.39000   0.41892 
 0.40000   0.40896 
 0.41000   0.39852 
 0.42000   0.38754 
 0.43000   0.37594 
 0.44000   0.36365 
 0.45000   0.35054 
 0.46000   0.33649 
 0.47000   0.32134 
 0.48000   0.30486 
 0.49000   0.28677 
 0.50000   0.26667 
 0.51000   0.24397 
 0.52000   0.21776 
 0.53000   0.18642 
 0.54000   0.14650 
 0.55000   0.08647 
 0.55100   0.07771
 0.55200   0.06776
 0.55300   0.05600
 0.55400   0.04091
 0.55514   0.00000 
/
\plot
 0.98146   0.00000
 0.98200   0.09065
 0.98300   0.15109
 0.98400   0.19192
 0.98500   0.22417
 0.98600   0.25125
 0.98700   0.27471
 0.98800   0.29548
 0.98900   0.31412
 0.99000   0.33102 
 1.00000   0.44444 
 1.01000   0.50686 
 1.02000   0.54527 
 1.03000   0.56991 
 1.04000   0.58585 
 1.05000   0.59600 
 1.06000   0.60211 
 1.07000   0.60534 
 1.08000   0.60646 
 1.09000   0.60600 
 1.10000   0.60436 
 1.11000   0.60179 
 1.12000   0.59852 
 1.13000   0.59468 
 1.14000   0.59039 
 1.15000   0.58574 
 1.16000   0.58079 
 1.17000   0.57559 
 1.18000   0.57018 
 1.19000   0.56459 
 1.20000   0.55884 
 1.21000   0.55294 
 1.22000   0.54690 
 1.23000   0.54073 
 1.24000   0.53443 
 1.25000   0.52799 
 1.26000   0.52143 
 1.27000   0.51473 
 1.28000   0.50789 
 1.29000   0.50089 
 1.30000   0.49373 
 1.31000   0.48639 
 1.32000   0.47886 
 1.33000   0.47112 
 1.34000   0.46314 
 1.35000   0.45492 
 1.36000   0.44642 
 1.37000   0.43761 
 1.38000   0.42845 
 1.39000   0.41892 
 1.40000   0.40896 
 1.41000   0.39852 
 1.42000   0.38754 
 1.43000   0.37594 
 1.44000   0.36365 
 1.45000   0.35054 
 1.46000   0.33649 
 1.47000   0.32134 
 1.48000   0.30486 
 1.49000   0.28677 
 1.50000   0.26667 
 1.51000   0.24397 
 1.52000   0.21776 
 1.53000   0.18642 
 1.54000   0.14650 
 1.55000   0.08647 
 1.55100   0.07771
 1.55200   0.06776
 1.55300   0.05600
 1.55400   0.04091
 1.55514   0.00000 
/
\plot
 1.98146   0.00000
 1.98200   0.09065
 1.98300   0.15109
 1.98400   0.19192
 1.98500   0.22417
 1.98600   0.25125
 1.98700   0.27471
 1.98800   0.29548
 1.98900   0.31412
 1.99000   0.33102 
 2.00000   0.44444 
/
\setsolid
   \setquadratic 
\plot
 0.00000   0.50000 
 0.01000   0.54514 
 0.02000   0.57634 
 0.03000   0.59808 
 0.04000   0.61303 
 0.05000   0.62293 
 0.06000   0.62893 
 0.07000   0.63188 
 0.08000   0.63237 
 0.09000   0.63086 
 0.10000   0.62768 
 0.11000   0.62307 
 0.12000   0.61724 
 0.13000   0.61034 
 0.14000   0.60247 
 0.15000   0.59373 
 0.16000   0.58416 
 0.17000   0.57382 
 0.18000   0.56272 
 0.19000   0.55087 
 0.20000   0.53827 
 0.21000   0.52490 
 0.22000   0.51072 
 0.23000   0.49568 
 0.24000   0.47970 
 0.25000   0.46271 
 0.26000   0.44458 
 0.27000   0.42516 
 0.28000   0.40425 
 0.29000   0.38159 
 0.30000   0.35681 
 0.31000   0.32941 
 0.32000   0.29862 
 0.33000   0.26319 
 0.34000   0.22081 
 0.35000   0.16614 
 0.36000   0.07586 
 0.36100   0.05941
 0.36200   0.03603
 0.36229   0.01801
 0.36258   0.00000
/
\plot
 0.97075   0.00000
 0.97100   0.05584
 0.97200   0.12462
 0.97300   0.16608
 0.97400   0.19821
 0.97500   0.22506
 0.97600   0.24838
 0.97700   0.26910
 0.97800   0.28780
 0.97900   0.30487
 0.98000   0.32058 
 0.99000   0.43233 
 1.00000   0.50000 
 1.01000   0.54514 
 1.02000   0.57634 
 1.03000   0.59808 
 1.04000   0.61303 
 1.05000   0.62293 
 1.06000   0.62893 
 1.07000   0.63188 
 1.08000   0.63237 
 1.09000   0.63086 
 1.10000   0.62768 
 1.11000   0.62307 
 1.12000   0.61724 
 1.13000   0.61034 
 1.14000   0.60247 
 1.15000   0.59373 
 1.16000   0.58416 
 1.17000   0.57382 
 1.18000   0.56272 
 1.19000   0.55087 
 1.20000   0.53827 
 1.21000   0.52490 
 1.22000   0.51072 
 1.23000   0.49568 
 1.24000   0.47970 
 1.25000   0.46271 
 1.26000   0.44458 
 1.27000   0.42516 
 1.28000   0.40425 
 1.29000   0.38159 
 1.30000   0.35681 
 1.31000   0.32941 
 1.32000   0.29862 
 1.33000   0.26319 
 1.34000   0.22081 
 1.35000   0.16614 
 1.36000   0.07586 
 1.36100   0.05941
 1.36200   0.03603
 1.36229   0.01801
 1.36258   0.00000
/
\plot
 1.97075   0.00000
 1.97100   0.05584
 1.97200   0.12462
 1.97300   0.16608
 1.97400   0.19821
 1.97500   0.22506
 1.97600   0.24838
 1.97700   0.26910
 1.97800   0.28780
 1.97900   0.30487
 1.98000   0.32058 
 1.99000   0.43233 
 2.00000   0.50000 
/
\put {2}   at 0.27 0.27
\put {1.5} at 0.40 0.31
\put {1}   at 0.68 0.36
\put {0.5} at 0.84 0.42
\put {2}   at 1.27 0.27
\put {1.5} at 1.40 0.31
\put {1}   at 1.68 0.36
\put {0.5} at 1.84 0.42
\endpicture
}

%
%

\newpage

\vspace*{20mm}
$$\hspace{-8mm}\plota\plotb$$
$$\hspace{-8mm}\plotc\plotd$$
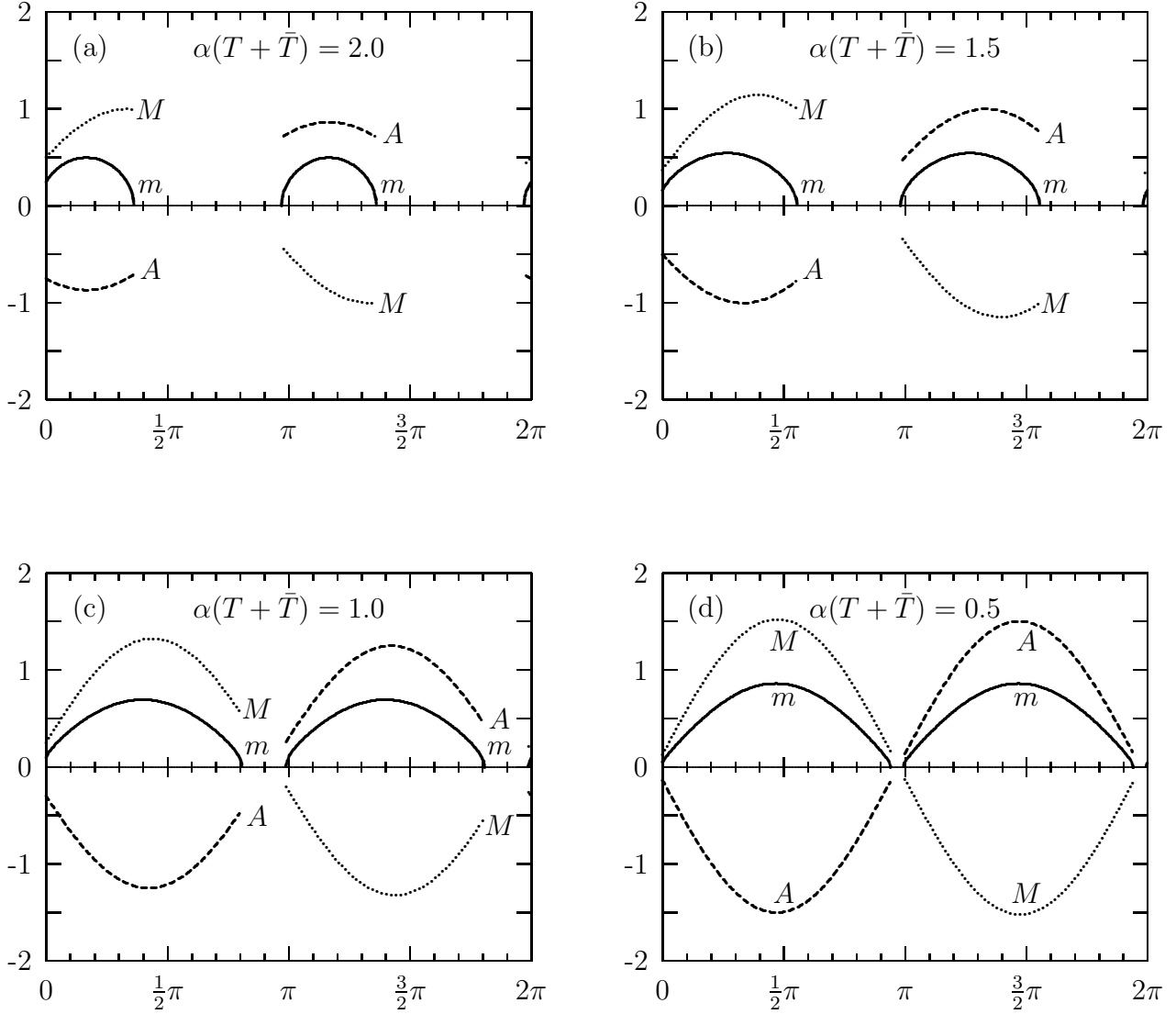
Figure~1: Soft parameters in unit of $m_{3/2}$ versus $\theta$
for different values of $\alpha(T+\bar T)$ in the $M$--theory limit.
Here $M$, $m$ and $A$ are the gaugino mass, the scalar mass and the trilinear
parameter respectively.

\newpage

$$\plotw$$
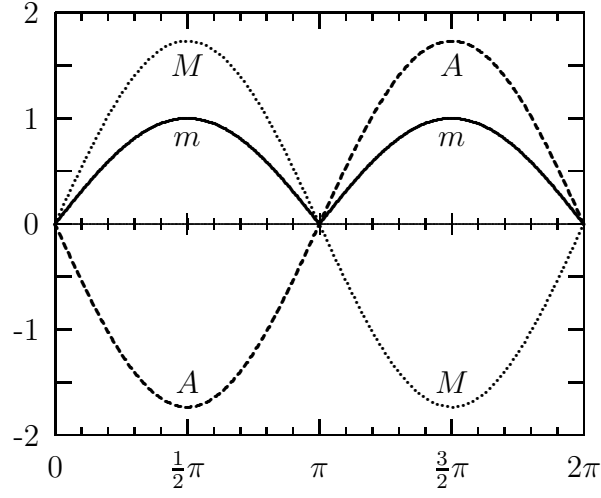
Figure~2: Soft parameters in unit of $m_{3/2}$ versus $\theta$
in the weakly coupled heterotic string limit.

\bigskip

$$\ratioplot$$
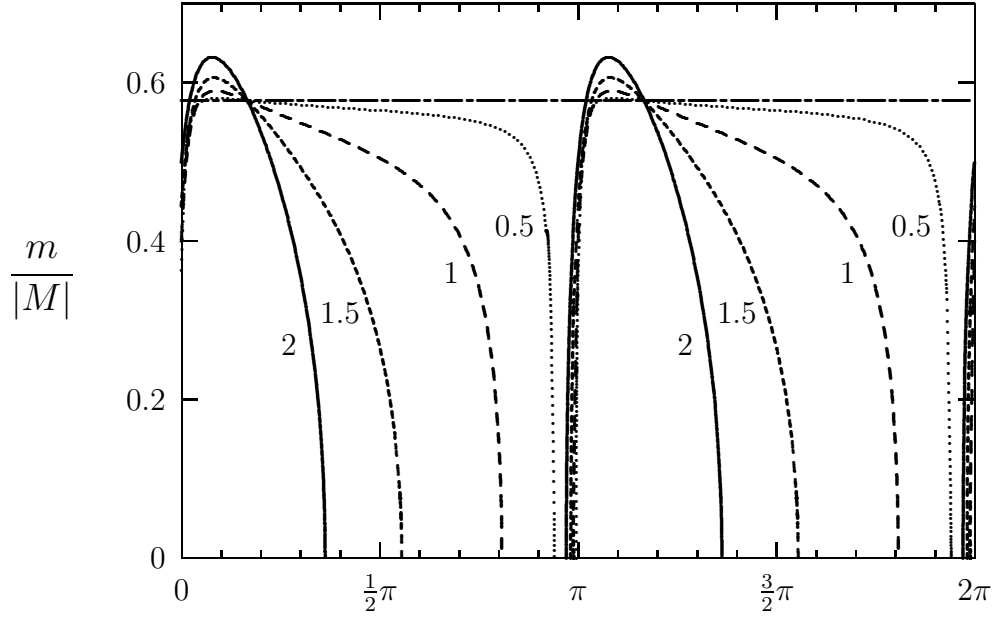
Figure~3: $m/|M|$ versus $\theta$ for different values of $\alpha(T+\bar T)$.
The straight line corresponds to the weakly coupled heterotic string limit.

\end{document}